\documentclass[journal, preprint]{IEEEtran}
\usepackage[hidelinks]{hyperref}
\usepackage{graphicx, import}
\usepackage{textcomp, gensymb}
\usepackage{amsmath}
\usepackage{epstopdf}
\usepackage{textcase}
\usepackage{multirow} 
\usepackage{amsfonts}
\usepackage{subcaption}
\usepackage{gensymb}
\usepackage{color,soul}
\allowdisplaybreaks

\begin{document}

\title{Robust EMRAN-aided Coupled Controller for Autonomous Vehicles}
\author{Sauranil Debarshi, Suresh Sundaram, and Narasimhan Sundararajan%
\thanks{Sauranil Debarshi and Suresh Sundaram are with the Department of Aerospace Engineering, Indian Institute of Science, Bangalore, India. (email: \href{mailto: sauranild@iisc.ac.in}{sauranild@iisc.ac.in}, \href{mailto: vssuresh@iisc.ac.in}{vssuresh@iisc.ac.in})}%
\thanks{Narasimhan Sundararajan is a Technical Consultant at the WIRIN Lab, Indian Institute of Science, Bangalore, India.  (email: \href{mailto: ensundara@gmail.com}{ensundara@gmail.com})}}

\maketitle

\begin{abstract}
This paper presents a coupled, neural network-aided longitudinal cruise and lateral path-tracking controller for an autonomous vehicle with model uncertainties and experiencing unknown external disturbances. Using a feedback error learning mechanism, an inverse vehicle dynamics learning scheme utilizing an adaptive Radial Basis Function (RBF) neural network, referred to as the Extended Minimal Resource Allocating Network (EMRAN) is employed. EMRAN uses an extended Kalman filter for online learning and weight updates, and also incorporates a growing/pruning strategy for maintaining a compact network for easier real-time implementation. The online learning algorithm handles the parametric uncertainties and eliminates the effect of unknown disturbances on the road. Combined with a self-regulating learning scheme for improving generalization performance, the proposed EMRAN-aided control architecture aids a basic PID cruise and Stanley path-tracking controllers in a coupled form. Its performance and robustness to various disturbances and uncertainties are compared with the conventional PID and Stanley controllers, along with a comparison with a fuzzy-based PID controller and an active disturbance rejection control (ADRC) scheme. Simulation results are presented for both slow and high speed scenarios. The root mean square (RMS) and maximum tracking errors clearly indicate the effectiveness of the proposed control scheme in achieving better tracking performance in autonomous vehicles under unknown environments.
\end{abstract}


\section{Introduction}

\IEEEPARstart{R}{ecent} technological developments in Advanced Driver Assistance Systems (ADAS) such as adaptive cruise control, lane keeping assistance, and automated parking have opened the doors for different prototypes of autonomous vehicles (AVs) to operate on the road. With the number of vehicle fatalities on the rise due to traffic congestion, human error, and lack of safety features, AVs are increasingly gaining attention as solutions to the above problems and for improving general road safety\unskip~\cite{eskandarian2012handbook}. Smart mobility features for perception, planning, control, and situational awareness on-board these vehicles\unskip~\cite{1131305:22465598} allow them to safely navigate without any human operator through complex and unstructured environments. One inherent attribute that makes an AV possess such a high level of intelligence is its ability to simultaneously control both the longitudinal and lateral dynamics while ensuring stability and ride comfort.

In autonomous driving, longitudinal control methods are designed to control the speed using the throttle and brake, while a lateral controller automatically steers the vehicle along a reference trajectory\unskip~\cite{1131305:22465612}. The longitudinal controller maintains a constant speed and keeps a safe distance behind another vehicle, using methods like Adaptive Cruise Control (ACC)\unskip~\cite{1131305:22465613}, emergency brake assist system\unskip~\cite{1131305:22465614}, and car-following\unskip~\cite{chen2016genetic}. Several control algorithms utilizing Model Predictive Control (MPC)\unskip~\cite{shakouri2014nonlinear}, extremum seeking approach\unskip~\cite{dinccmen2018emergency}, proportional-integral (PI) control\unskip~\cite{9560150}, and model reference adaptive control\unskip~\cite{raffin2017adaptive} have been proposed in the literature to control the longitudinal dynamics. On the contrary, the focus of the lateral control system is to avoid obstacles and assist in lane-keeping and lane-changing maneuvers. The authors in\unskip~\cite{8984590} surveyed some of the current state-of-the-art lateral control methods based on adaptive PID, fuzzy logic, and neural networks. Other steering algorithms present in the literature include a lateral $\text{H}_\infty $ controller\unskip~\cite{1131305:22465629}, linear quadratic regulator (LQR)\unskip~\cite{1131305:22465627}, and a combination of backstepping and sliding mode controllers (SMCs)\unskip~\cite{1131305:22465625}.

Recently, use of learning-based controllers is increasingly becoming popular in AVs because of their ability to self-optimize and adapt to the dynamics of the environment. Neural controllers, for instance, have been widely used in the literature for AV control. Taghavifar \textit{et al.}\unskip~\cite{1131305:22465616} proposed a type-2 fuzzy neural PID controller to minimize the heading and lateral errors during path-tracking. Additionally, they developed an extended Kalman filter-based adaptive observer to eliminate the effects of external disturbances and parametric uncertainties. A backstepping variable structure control coupled to a RBF neural network was presented in\unskip~\cite{1131305:22465617}. The BVSC steers the vehicle, while the RBFNN eliminates the errors by acting as the estimator for the tire nonlinearities. In another study\unskip~\cite{1131305:22465609}, a SMC with an RBF neural network was developed for improving the tracking of the speed. The RBF network with its adaptive and universal function approximation ability reduces the large errors of the SMC and improves the robustness of the system.  More recently, a comprehensive study on a wide range of longitudinal and lateral vehicle control methods based on deep learning was conducted in\unskip~\cite{8951131}. Various supervised and reinforcement learning strategies were compared in terms of network complexity, learning capabilities, and performance. It has been concluded that while deep learning has shown great promise in AV control applications, it also presents numerous challenges for actual deployment in vehicles. The need for large amount of data that captures all possible driving scenarios, computational effort, and selection of network architecture limits the use of deep learning-based control in AVs.

It should be noted that much of the existing research addresses the longitudinal and lateral controllers separately, assuming that there is no dynamic interaction (coupling) between them. However, it is not the case in an actual vehicle, and in many critical autonomous driving scenarios, a coupled control strategy is required\unskip~\cite{amer2017modelling}. A simultaneous longitudinal and lateral control architecture was presented in\unskip~\cite{attia2014combined}. A nonlinear MPC was employed as the steering controller and a Lyapunov-based control law considering the powertrain dynamics was used for the longitudinal speed-tracking. Another coupled method was shown in\unskip~\cite{8570020} by quantitatively comparing two different deep learning models, a Multi-layer Perceptron (MLP) and a Convolutional Neural Network (CNN). The CNN proved to be a better controller in terms of accuracy and smoothness of input commands. A recent study by Tork \textit{et al.}\unskip~\cite{TORK2021126} describes an integrated control system using an adaptive multi-layer neural network based on an interval type-2 fuzzy activation function. They performed Double Lane Change (DLC) maneuvers with and without parametric uncertainties, and their proposed controller outperformed other neural network-based controllers. However, effect of external disturbances such as slippery road conditions and different velocities were not taken into account in their study.

In this paper, we propose a novel neuro-aided coupled longitudinal and lateral control scheme for an AV by utilizing an inverse vehicle dynamics learning technique. We employ a previously developed RBF neural network referred to as the Extended Minimal Resource Allocating Network (EMRAN)\unskip~\cite{1131305:22465619} and integrate a self-regulated learning mechanism\unskip~\cite{suresh2010sequential} as an extension for efficient training with fewer samples. EMRAN is a fast, adaptive, sequential learning algorithm that requires no a priori training and overcomes the challenges of deep learning-based controllers discussed earlier. It starts with zero hidden neurons and incorporates a growing/pruning strategy, making it a computationally inexpensive learning algorithm with a compact and efficient network suitable for real-time implementation in AVs. To the best of the authors' knowledge, no prior studies have examined an online inverse vehicle dynamics learning model for developing an integrated control system that is robust to unknown disturbances and uncertainties. Another major drawback of most of the learning-based approaches such as Deep Neural Networks (DNNs), is that they scale poorly and provide no generalization guarantee\unskip~\cite{8951131}. The online inverse dynamics learning method addresses this issue by allowing a neural controller to be used in conjunction with any feedback controller for improving the tracking performance. The main contributions of the paper are as follows: i) A model-free, coupled EMRAN-aided controller is proposed for longitudinal cruise control and lateral path-tracking. A conventional PID aided by an EMRAN neural network (PID-EMRAN) provides the acceleration/deceleration commands for maintaining the desired velocity. Another EMRAN is used to aid a Stanley\unskip~\cite{1131305:22465623} controller (Stanley-EMRAN) for steering the vehicle along a reference trajectory; ii) The EMRAN-aided controller supports online adaptability and learns to approximate the inverse dynamics of the vehicle using a feedback error learning strategy. It ensures stability and provides robustness against various unknown disturbances and parametric uncertainties; iii) The neuro-aided control architecture could be used as an add-on with any feedback controller for improving the tracking performance, thereby providing flexibility in the overall controller design.

The performance of the EMRAN-aided controller has been evaluated against conventional PID and Stanley methods, for both coupled and decoupled states. Additionally, the proposed controller is compared with a recently developed type-2 Fuzzy PID controller\unskip~\cite{1131305:22465616} and an active disturbance rejection control scheme\unskip~\cite{1131305:22465637}. The obtained results highlight the benefits of using EMRAN for real-time longitudinal and lateral control of AVs.

Rest of the paper is organized as follows: In Section II, the problem statement related to AV control is formulated. Section III discusses the EMRAN-aided inverse dynamics learning controller in detail. Performance evaluation using simulation results is shown in Section IV. Finally, the conclusions based on the study are summarized in Section V.

\bgroup
\begin{figure}[!t]
\centering \makeatletter\IfFileExists{images/fig1.eps}{\includegraphics[width=\columnwidth]{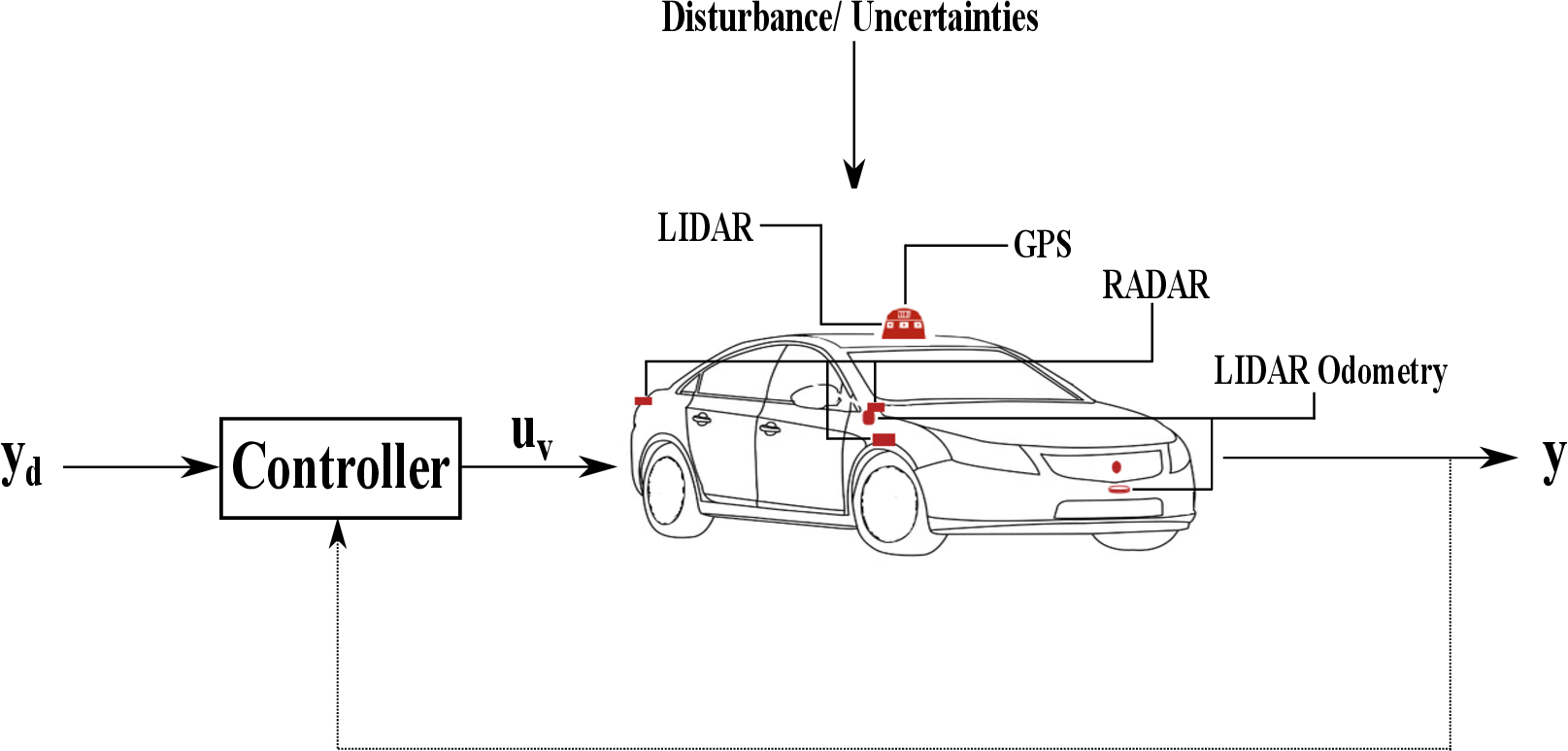}}{}
\makeatother 
\caption{{Representation of a typical vehicle control architecture.}}
\label{f-7c7cf8ba592a}
\end{figure}
\egroup

\section{Problem Formulation}
Before presenting the mathematical model of the nonlinear vehicle used for the simulations, we first formulate the control problem in AVs to track a reference signal $\mathbf{y_d} = \{v_r, \{y_r, \psi_r\}\}$, where, $v_r$ represents the reference cruise speed, and $y_r$ and $\psi_r$ respectively denote the reference lateral displacement and desired yaw angle of the vehicle. These reference signals are generated by a motion planning algorithm in an actual AV and a discussion on the same is outside the scope of the current study. As shown in Fig.~\ref{f-7c7cf8ba592a}, an AV requires multiple onboard sensors to replicate the human understanding of its location, perception, and navigation. The sensors provide critical information to the controller for generating control inputs such that the vehicle can track the reference signal with minimum errors. In this paper, both the longitudinal and lateral control problem in AVs are investigated through cruise control and DLC maneuvers. 

The state-space model of the vehicle at time $t$ is given by:
\begin{equation}
    \sum \dot{\mathbf{x}} = g(\mathbf{x}, \mathbf{u_{v}}, t)
\tag{1}\label{dfg-33598e9dae38}\end{equation}
where the dynamics of the vehicle is denoted by the nonlinear function $g()$. $\mathbf{u_v} \in \mathbb{R}^q $ is the control input and $\mathbf{x} \in \mathbb{R}^m $ is the state vector. Also, $\mathbf{u_v} $ belongs to a class of permissible inputs, given by:
\begin{equation}
    U~:= \left\{\mathbf{u_v}: ||\mathbf{u_v}(t)|| \leq \zeta, t > t_0 \right\}
\tag{2}\label{dfg-a59c7f5841b4}\end{equation}
where $\zeta $ is a real positive number and it puts a constraint on the control input such that the vehicle follows the reference signals as closely as possible after a certain time instant, $t_0$, and without losing stability. With the constraint on $\mathbf{u_v} $, we define the control objective as:
\begin{equation}
||\mathbf{y_d} - \mathbf{y}|| \to 0
\tag{3}\label{dfg-12e413f00143}\end{equation}
where $\mathbf{y_d} $ is the reference signal and $\mathbf{y} $ is the output of the vehicle. 

A nonlinear single-track or ``bicycle model'' is used for the simulations, as depicted in Fig.~\ref{f-fd63b1b6f1a0}. Due to the symmetry of the left and right side of a vehicle, the bicycle model provides a simplified, and at the same time, an accurate representation of the vehicle’s dynamics.

\bgroup
\begin{figure}[!t]
\centering \makeatletter\IfFileExists{images/fig2.eps}{\includegraphics[width=\columnwidth]{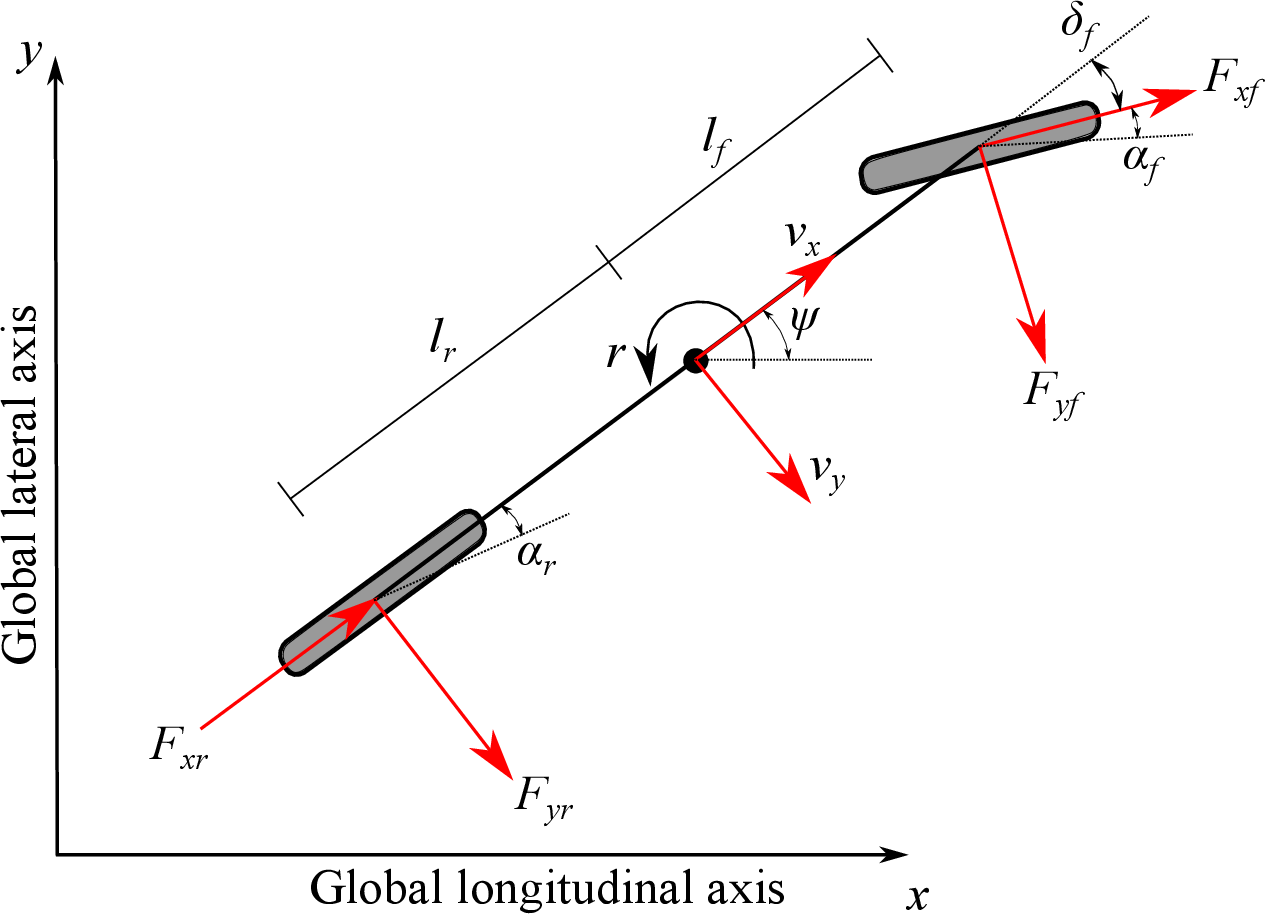}}{}
\makeatother 
\caption{{Nonlinear single-track vehicle dynamics model.}}
\label{f-fd63b1b6f1a0}
\end{figure}
\egroup
In \unskip~\cite{1131305:22465645} and\unskip~\cite{1131305:22465644}, the dynamics of the bicycle model are described, with the longitudinal and lateral velocities in the global frame given by:
\begin{equation}
    \begin{aligned}
    &\dot{x} = v_x\cos{\psi} - v_y\sin{\psi} 
    \\
    &\dot{y} = v_x\sin{\psi} + v_y\cos{\psi} 
    \end{aligned}
    \tag{4}\label{Eq: vel}
\end{equation}
where, $v_x $ and $v_y $ represent the longitudinal and lateral velocities in the vehicle frame respectively and $\psi $ is the yaw angle, whose rate and acceleration are defined as:
\begin{equation}
    \begin{aligned}
    &\dot{\psi} = r 
    \\
    &\dot{r} = \dfrac{1}{I_z}(F_{yf}l_f - F_{yr}l_r) 
    \end{aligned}
    \tag{5}
\end{equation}
where, $F_{yf} $ is the lateral tire force of the front wheel and $F_{yr} $ is the lateral tire force of the rear wheel. $l_f $ and $l_r $ are respectively the distances to the front and rear axles from the center of gravity (CoG) of the vehicle, while $I_z $ is the yaw moment of inertia. 

The vehicle's longitudinal and lateral acceleration are obtained using:
\begin{equation}
    \begin{aligned}
    &\dot{v_x} = \dfrac{1}{m}(F_{xf} + F_{xr}) + v_y r 
    \\
    &\dot{v_y} = \dfrac{1}{m}(F_{yf} + F_{yr}) - v_x r
    \end{aligned}
    \tag{6}\label{Eq: acc}
\end{equation}
where, $F_{xf} $ and $F_{xr} $ denote the longitudinal tire forces of the front and rear wheels respectively, $m $ is the mass of the vehicle, and $r $ is the yaw rate. 

The wheels' angular velocities are calculated from the drive ($T_d $) and brake ($T_b $) torques as: 
\begin{equation}
\begin{aligned}
    &\omega_{wf} = \dfrac{1}{I_w}(T_d - R_tF_{xf} - T_b)  
    \\
    &\omega_{wr} = \dfrac{1}{I_w}(T_d - R_tF_{xr} - T_b) 
\end{aligned}
\tag{7}\label{dfg-b4a622b77fc7}\end{equation}
where, $R_t $ is the static radius of the wheel with moment of inertia $I_w $.

For estimating the forces on the wheels in the longitudinal and lateral directions, an analytical tire model developed by Salaani\unskip~\cite{1131305:22465624} has been used. This nonlinear tire model captures the wheel-road dynamics accurately and has been validated from experimental data using four different tires. The longitudinal and lateral wheel forces are given as:
\begin{align}
\begin{split}
    F_{xi} &= -\dfrac{\kappa_i F_{zi} C_{\kappa i} f_{ai}(\sigma )}{\sqrt{\left (\dfrac{C_{\kappa i} \kappa_i}{\mu_{pxi}} \right)^{2} + \left(\dfrac{C_{\alpha i} \tan(\alpha_i)}{\mu_{pyi}}\right)^{2}}} \\
    &\quad -\dfrac{\kappa_i F_{zi} C_{\kappa si} f_{si}(\sigma )}{\sqrt{\left(\dfrac{C_{\kappa si} \kappa_i}{\mu_{sxi}}\right)^{2} + \left(\dfrac{C_{\alpha i} \tan(\alpha_i)}{\mu_{syi}}\right)^{2}}} \qquad (i = f, r) \\
    F_{yi} &= \dfrac{C_{\alpha i} \tan(\alpha_i) F_{zi} f_{ai}(\sigma )}{\sqrt{\left(\dfrac{C_{\kappa i} \kappa_i}{\mu_{pxi}} \right)^{2} + \left(\dfrac{C_{\alpha i} \tan(\alpha_i)}{\mu_{pyi}}\right)^{2}}} \\
    &\quad +\dfrac{C_{\alpha i} \tan(\alpha_i) F_{zi} f_{si}(\sigma )}{\sqrt{\left(\dfrac{C_{\kappa si} \kappa_i}{\mu_{sxi}} \right)^{2} + \left(\dfrac{C_{\alpha i} \tan(\alpha_i)}{\mu_{syi}}\right)^{2}}}
    \qquad (i = f, r)
\end{split}
\tag{8}\label{dfg-5092d28933e9}\end{align}
where $F_{zi} $, $\kappa_i $, and $\alpha_i $ are the wheel load, longitudinal slip ratio, and slip angle, respectively. $C_{\kappa i} $ and $C_{\alpha i} $ denote the longitudinal and lateral tire stiffness, $f_{ai}() $ and $f_{si}() $ are the adhesion and sliding functions, $\sigma $ is the adhesion potential rate, and lastly $C_{\kappa si} $ represents the longitudinal stiffness in sliding mode. The peak/sliding coefficients of friction in the longitudinal and lateral directions are given by $\mu_{pxi} $, $\mu_{sxi} $, $\mu_{pyi} $, and $\mu_{syi} $, where the subscript $i $ denotes the front and rear tires.

Using small angle approximations, the tire slip angle in the front ($\alpha_f $) and rear ($\alpha_r $) can be linearized\unskip~\cite{1131305:22465646} as:
\begin{equation}
    \begin{aligned}
    &\alpha_f =  \arctan\bigg(\dfrac{\dot{y} + l_f\dot{\psi}}{\dot{x}}\bigg) - \delta_f \approx \dfrac{\dot{y} + l_f\dot{\psi}}{\dot{x}} - \delta_f \\
    &\alpha_r =  \arctan\bigg(\dfrac{\dot{y} - l_r\dot{\psi}}{\dot{x}}\bigg) \approx \dfrac{\dot{y} - l_r\dot{\psi}}{\dot{x}}
    \end{aligned}
\tag{9}\label{dfg-d28dda6818d0}\end{equation}
where $\delta_f $ represents the steering angle. Additionally, the tire slip ratio ($\kappa_i $)\unskip~\cite{1131305:22465647} is defined as:
\begin{equation}
    \kappa_i = \dfrac{R_{ei}\omega_{wi} - v_x}{\max(R_{ei}\omega_{wi}, v_x, \epsilon)} \qquad (i = f, r)
\tag{10}\label{dfg-f55acc209d6e}\end{equation}
where $R_{ei} $ is the effective tire radius and $\epsilon $ is a small constant ($\epsilon \ll 1 $) to avoid zero denominator.

Next, we describe the online inverse dynamics learning controller for AVs in detail.
    
\section{Inverse Dynamics Learning EMRAN Controller}
In this section, the EMRAN-aided coupled longitudinal and lateral controller is presented. An inverse dynamics learning with self-regulation is proposed to achieve the tracking objective. The working of EMRAN is briefly discussed next.

\subsection{Extended Minimal Resource Allocating Network (EMRAN)}EMRAN is a fast, online learning algorithm developed by Li \textit{et al.}\unskip~\cite{1131305:22465619}, ideal for real-time application. It implements a compact RBF neural network by incorporating a fully adaptive learning strategy, with the capability to add and prune the hidden neurons based on the network inputs and the responses of the controlled object. It is an extension of the previously developed sequential learning RBF algorithm called the Minimal Resource Allocation Network (MRAN)\unskip~\cite{6796339, 661125}. MRAN also utilizes a growing/pruning strategy for ensuring compactness. However, unlike EMRAN in which the parameters of only the nearest hidden neuron (in a norm sense) are updated, MRAN updates the centers, widths and weights of all the hidden neurons at every timestep. This causes the size of the matrices to be updated to become large as the hidden neurons increase and the RBF network structure becomes more complex computationally, thereby limiting the use of MRAN for real-time implementations.

\bgroup
\begin{figure}[!t]
\centering \makeatletter\IfFileExists{images/fig3.eps}{\includegraphics[width=\columnwidth]{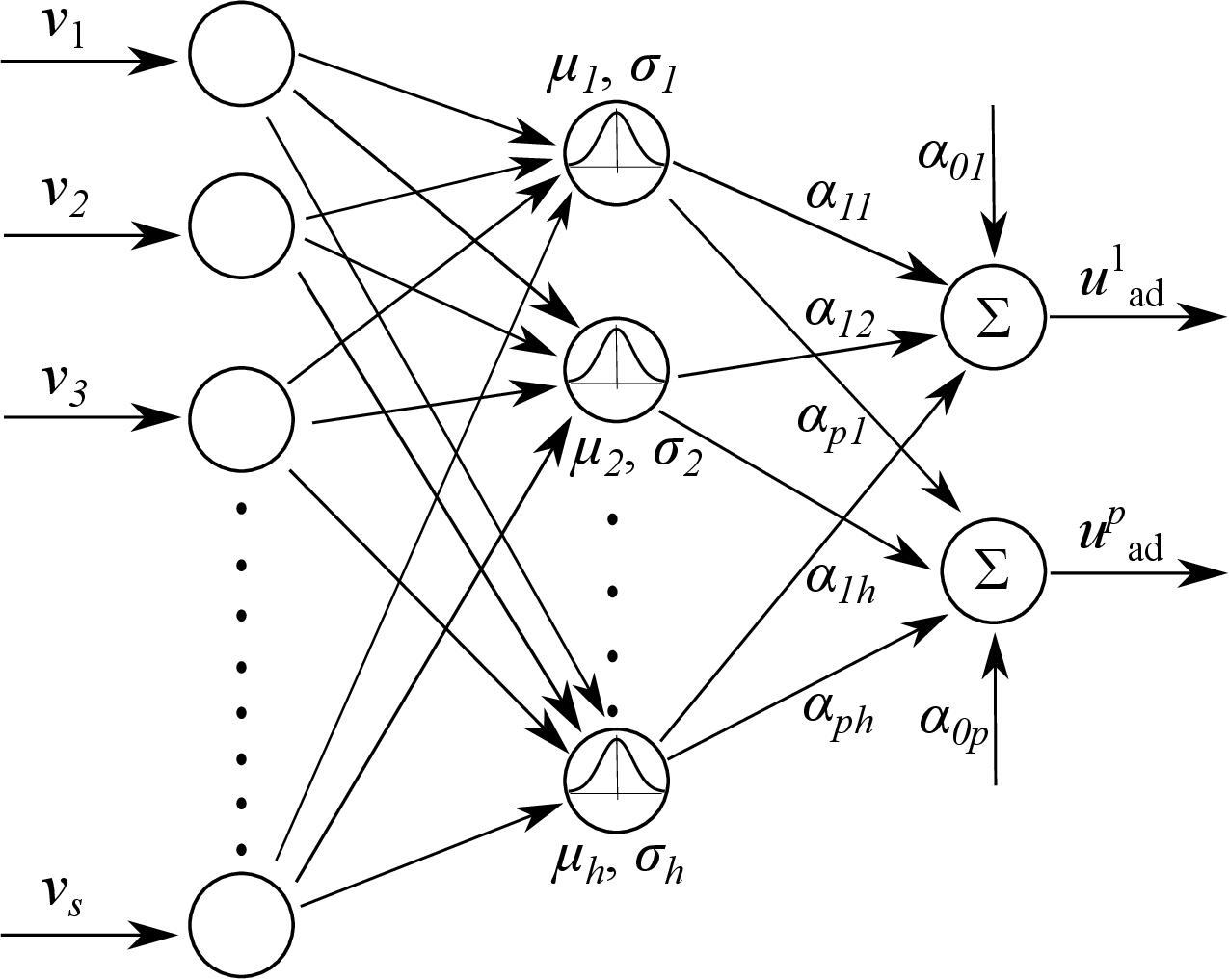}}{}
\makeatother 
\caption{{EMRAN architecture with Gaussian activation functions.}}
\label{f-9b40bfc4092e}
\end{figure}
\egroup
Figure~\ref{f-9b40bfc4092e} shows the EMRAN architecture consisting of a single hidden layer with $h $ hidden neurons. It initially starts with zero hidden neurons and they are added/pruned based on a heuristic condition. The coefficients $\alpha_{jk} $ are the interconnection weights between the hidden and the output layer. The activation functions used in the hidden layer are Gaussian, whose outputs are given by:
\begin{equation}
    z^{k} = exp^{\tfrac{-||\textbf{v} - \boldsymbol{\mu}_k||^{2}}{2(\boldsymbol{\sigma }_k)^{2}}} \qquad (k = 1, 2,....,h)
\tag{11}\label{dfg-e9997e0d0c7f}\end{equation}
where $\mathbf v \in \mathbb{R}^s$ is the network input, while $\boldsymbol{\mu}_k, \boldsymbol{\sigma }_k \in \mathbb{R} $ are the center and width of the hidden neurons respectively. Gaussian activation functions have good local interpolation (each neuron responds only to a specified region of the input space) and global approximation ability\unskip~\cite{6796339, 1131305:22465619}. EMRAN, thus, explicitly stores information regarding the input characteristics, instead of merely using the information for updating the network parameters.  The outputs of the neural network are then given by:
\begin{equation}
    u^{j}_{ad} = \sum_{k = 1}^{h}\alpha_{jk}z^{k} + \alpha_{\textit{0}j} \qquad (j = 1, 2,....,p)
\tag{12}\label{dfg-d7da65f399b2}\end{equation}
where $\alpha_{\textit{0}j} $ are the biases at the output layer and $p $ is the total number of outputs of the network. 

\begin{figure*}[!t]
    \centering
    \begin{subfigure}[t]{0.48\textwidth}
        \centering
        \includegraphics[width=\textwidth]{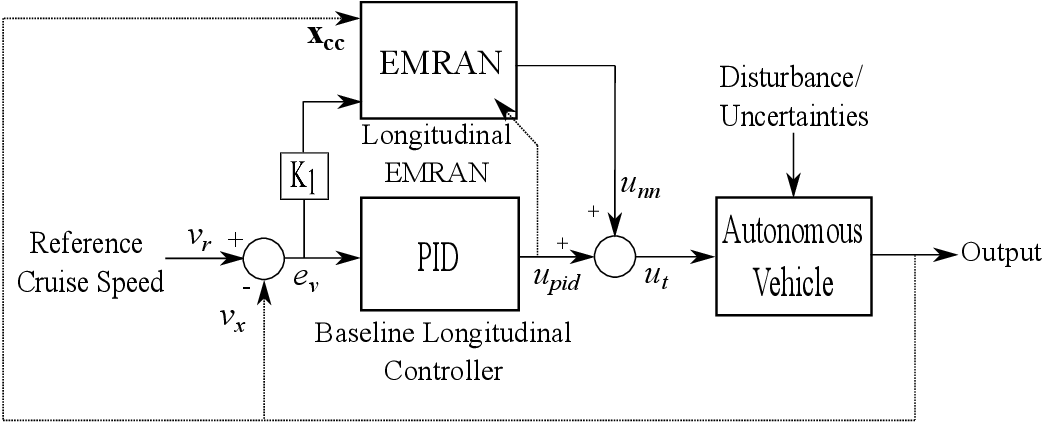}
        \caption{Longitudinal cruise control subsystem.}
        \label{fig-4a}
    \end{subfigure}%
    ~ 
    \begin{subfigure}[t]{0.48\textwidth}
        \centering
        \includegraphics[width=\textwidth]{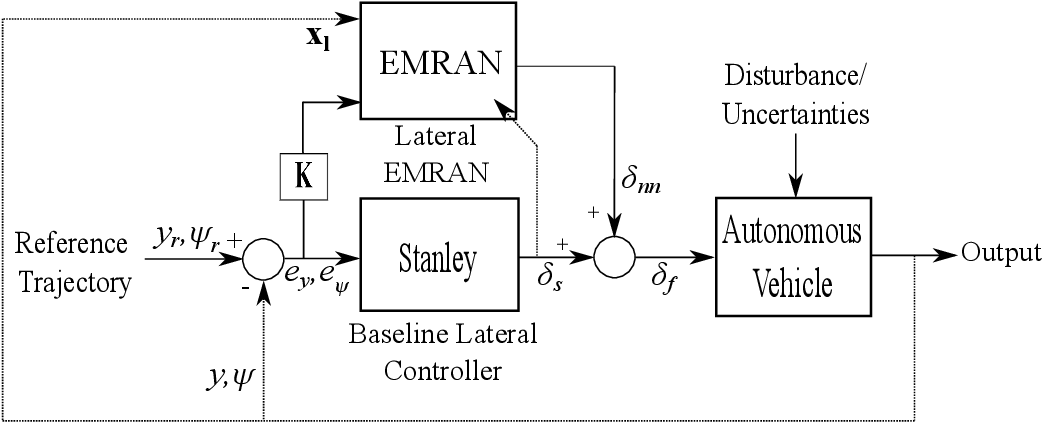}
        \caption{Lateral path-tracking subsystem.}
        \label{fig-4b}
    \end{subfigure}
    \caption{Schematic of the EMRAN-aided longitudinal and lateral control architectures.}
    \label{fig-4}
\end{figure*}

EMRAN starts with no hidden neurons. A new neuron is added when the following criteria are satisfied at any time step, $\tau $:
\begin{equation}
    \begin{aligned}
    &||\textbf{v}[\tau] - \boldsymbol{\mu}_w[\tau]|| > \epsilon_1[\tau] \\
    &||\textbf{y}_e[\tau]||^{2} \ge \epsilon_2 \\
    &J_{rmse} = \sqrt{\dfrac{\sum\limits_{\tau = l - S_w + 1}^{l}||\textbf{y}_e[\tau]||^{2}}{S_w}} \ge \epsilon_3 
    \end{aligned}
\tag{13}\label{dfg-0f52e89d0736}\end{equation}
where $\epsilon_1[\tau] $ = max[$\epsilon_{max} $$\gamma^{\tau-1} $, $\epsilon_{min} $]. EMRAN starts with $\epsilon_1[\tau] = \epsilon_{max}$, the largest scale of interest, which is typically the size of the entire input space of nonzero probability density, and decays exponentially until it reaches $\epsilon_{min}$. $\gamma $ is the decay constant between 0 and 1 and represents the scale of resolution\unskip~\cite{1131305:22465620}. $\textbf{y}_e[\tau] $ is the error in the network output, $\boldsymbol{\mu}_w $ is the center of the hidden neuron closest to $\textbf{v}[\tau] $, $S_w $ is the sliding window's length and $\epsilon_1 $, $\epsilon_2 $, and $\epsilon_3 $ are the thresholds that needs to be selected appropriately. The distance between the new observation and all the existing nodes is compared by the first criterion. The second criterion ascertains that the existing neurons are sufficient to produce a reasonable network output. The third criterion is based on the root mean square error for the window of samples $S_w $, which controls the noise from over-fitting the neurons. The parameters of the newly added hidden neuron are given by:
\begin{equation}
    \begin{aligned}
    &\boldsymbol{\alpha}_{k+1}[\tau] = \mathbf{y}_e[\tau - 1] \\
    &\boldsymbol{\mu}_{k+1}[\tau] = \mathbf{v}[\tau] \\
    &\boldsymbol{\sigma }_{k+1}[\tau] = \kappa||\mathbf{v}[\tau] - \boldsymbol{\mu}_w[\tau]|| 
    \end{aligned}
\tag{14}\label{dfg-a6975b140731}\end{equation}
where $\kappa $ determines the overlap of the responses of the hidden neurons in the input space. When the above criteria are not satisfied, an Extended Kalman Filter (EKF) updates the parameters of the neuron whose center is nearest to the network input data $\textbf{v}$. This neuron is referred to as the ``winner neuron'', whose parameters at the $\tau$th instant are updated as:
\begin{equation}
    \textbf{NP}^{w}[\tau] = \textbf{NP}^{w}[\tau-1] + \textbf{K}^{w}[\tau]||\textbf{y}_e[\tau - 1]||
\tag{15}\label{dfg-7fd77e3f834b}\end{equation}
where $\textbf{K}^{w}[\tau] $ is the Kalman gain matrix given by:
\begin{equation}
    \textbf{K}^{w}[\tau]=\textbf{P}^{w}[\tau-1]\textbf{B}^{w}[\tau](\textbf{R}[\tau] + \textbf{B}^{w}[\tau]^{T} \textbf{P}^{w}[\tau-1]\textbf{B}^{w}[\tau])^{-1} 
\tag{16}\label{dfg-c28dec215d14}\end{equation}
where $\mathbf{B}^{w}[\tau] = \nabla_{\mathbf{w}} \mathbf{u}_{\text{ad}} $ is the gradient matrix of $\mathbf{u}_{\text{ad}} $ with respect to the parameter vector $\textbf{w}[\tau] $ evaluated at $\mathbf{w}[\tau - 1] $. $\textbf{R}[\tau] $ is the variance of the measurement noise and $\mathbf{P}^{w}[\tau] $ is the error of the covariance matrix, which is updated by:
\begin{equation}
    \textbf{P}^{w}[\tau] = (\textbf{I} - \textbf{K}^{w}[\tau]\textbf{B}^{w}[\tau]^{T})\textbf{P}^{w}[\tau-1] + q\textbf{I}
\tag{17}\label{dfg-3ff235f6b153}\end{equation}
where $q $ is a scalar quantity that determines the allowed random step in the direction of the gradient vector. When a new hidden unit is added to the network, the dimensionality of the covariance matrix increases to:
\begin{equation}
    \mathbf{P}^{w}[\tau] = 
    \begin{bmatrix}
    \mathbf{P}^{w}[\tau-1] & 0\\
    0 & P_0\mathbf{I}
    \end{bmatrix}
\tag{18}\label{dfg-46daa37a91d1}\end{equation}
where $P_0 $ is a scalar value that represents the uncertainty in the initial parameters of the new hidden neuron.

EMRAN incorporates a pruning strategy for maintaining a compact network. It ensures that the neurons that have not been contributing significantly (based on a threshold parameter ($\delta $)) to the network performance for a predefined period of time ($N_w $), are pruned from the network. In addition, two hidden neurons are combined into a single hidden neuron if they are found to be closer to one another, as defined by a threshold value. It results in a network that is computationally inexpensive and adapted to fast real-time online applications. Moreover, since only the parameters of the winner neuron are updated, the computations required to update its parameters is $O(l^{3}) $, where $l $ is the number of parameters of the nearest neuron. Therefore, the total computational burden at each step is $8^{3} $ floating point operations (FLOPS) (1 FLOP $= 1e^{-06} $ s), which is relatively small. Control output calculation is $O(h) $, where $h $ is the number of hidden neurons. 

To further improve the online learning process and avoid overtraining the EMRAN network, a self-regulated learning scheme\unskip~\cite{suresh2010sequential, savitha2012meta} has been incorporated. The EKF, instead of using every training samples for updating the parameters of the winner neuron, utilizes only a subset of the training data based on certain thresholds of the tracking and residual errors. It achieves similar control performance to a conventionally trained network while using fewer samples, allowing the EMRAN network to learn faster, with reduced computational effort. Next, the EMRAN-aided control architecture and its functioning for AV control are described.

\subsection{EMRAN-aided Inverse Vehicle Dynamics Learning}Consider the dynamics of the vehicle in (\ref{Eq: vel}-\ref{Eq: acc}) to be in the form of:
\begin{equation}
    \dot{\textbf{x}} = g(\textbf{x}, \mathbf{u_v})
\tag{19}\label{dfg-b5db189a1b06}\end{equation}
where $\mathbf{u_v} = [u_{t}, \delta_{f}] $ constitutes the control input vector to the AV, with $u_{t} $ representing the acceleration/deceleration command for longitudinal cruise control and $\delta_{f} $ denoting the steering angle for lateral path-tracking.  In this paper, we utilize the feedback error learning technique of Gomi and Kawato\unskip~\cite{1131305:22465622} to learn the inverse dynamics of the vehicle using the EMRAN neural network. By learning the inverse dynamics, EMRAN can compensate for the nonlinearities of the vehicle such that it follows the desired response set closely. The inversion can be represented by the equation:
\begin{equation}
    \mathbf{u_v} = g^{-1}(\dot{\mathbf{x}}, \mathbf{x})
\tag{20}\label{dfg-52a10733b3b9}\end{equation}
Further, if the function $g^{-1}() $ is changing with time due to external factors, the neural controllers can generate immediate corrective actions to compensate for such changes.

The architectures of the EMRAN-aided longitudinal and lateral control subsystems of the coupled controller are shown in Fig.~\ref{fig-4}, with their respective control input to the AV given by:
\begin{equation}
    \begin{aligned}
    &u_t = u_{pid} + u_{nn} \\
    &\delta_f = \delta_s + \delta_{nn}
    \end{aligned}
\tag{21}\label{dfg-743d217bf0ba}\end{equation}

The outputs of the baseline PID and Stanley controllers are denoted by a vector, $\textbf{u}_{\textbf{b}} = $ [$u_{pid} $, $\delta_s $] while their respective EMRAN outputs are represented by another vector as $\textbf{u}_\textbf{nn} = $ [$u_{nn} $, $\delta_{nn} $]. The outputs of the AV, namely, the velocity, lateral position and yaw rate are given as feedbacks to calculate their respective errors, which in turn form the inputs to the baseline controllers. In both the architectures, the baseline controllers provide the basic stability requirements and also generate the signals to train their EMRAN networks online. The EMRANs compensate for the disturbances and parametric uncertainties of the vehicle, thereby aiding the baseline controllers to achieve a better tracking performance. The online learning process of the proposed controller is discussed next. 

\begin{table}[!t]
\caption{{Vehicle Parameters.} }
\label{tw-1fba46a1b919}
\def\arraystretch{1}
\ignorespaces 
\centering 
\begin{tabular*}{0.7\columnwidth}{@{\extracolsep{\stretch{1}}}*{2}{c}@{}}
\hline Parameters & Values\\
\hline 
$m $ &
  1480 kg\\
$I_z $ &
  2350 kg.m\ensuremath{^{2}}\\
$l_f $ &
  1.05 m\\
$l_r $ &
  1.63 m\\
$C_f $ &
  67500 N/rad\\
$C_r $ &
  47500 N/rad\\
\hline 
\end{tabular*}\par 
\end{table}

\subsection{Online Learning Process}The inner loop of the longitudinal cruise control uses a fully tuned PID controller as the baseline controller, as shown in Fig.~\ref{fig-4a}. It generates acceleration and deceleration commands ($u_{pid} $) based on the differences between the reference ($v_r $) and actual ($v_x $) velocities. The drive torque ($T_d $) and brake torque ($T_b $) are ultimately computed from the acceleration/deceleration commands using dynamic equations. The input vector ($\textbf{x}_{\textbf{cc}} $) to the cruise control EMRAN consists of the longitudinal states of the vehicle, namely the position ($x $), velocity ($v_x $), and acceleration ($a_x $). Based on the postulate in\unskip~\cite{1131305:22465622}, the output of the PID controller ($u_{pid} = u_t - u_{nn}$) is used as the signal for updating the weights and neurons of the network. EMRAN learns the total control signal ($u_{nn} \rightarrow u_t$) over time and eventually driving the PID output to zero. This means that EMRAN has generated the inverse longitudinal dynamics of the vehicle and is using it for control. However, if EMRAN learns only from the baseline controller, there is the possibility that it will not achieve any better results than the PID, since the conventional controllers are not robust against unknown disturbances and uncertainties. Hence, to get a better performance and eliminate the effects of unwanted nonlinearities, the learning signal is modified by adding the output of the PID controller with a scaled ($\text{K}_1 $) velocity error signal. 

Similarly, for the lateral path-tracking subsystem, a Stanley controller\unskip~\cite{1131305:22465623} is used in the inner loop, as depicted in Fig.~\ref{fig-4b}. The Stanley controller is a geometrical path-tracking algorithm developed by the Stanford University's DARPA grand challenge team. It is a nonlinear feedback function of the cross-track/lateral ($e_y $) and heading ($e_{\psi} $) errors measured from the vehicle's front axle. The steering angle is given by:
\begin{equation}
    \delta_s = e_\psi + \tan^{-1} \left(\frac{k_fe_y}{v_x} \right)
\tag{22}\label{dfg-9ea7938c0b3e}\end{equation}
where $k_f $ is the position gain in the forward motion, and $v_x $ is the longitudinal velocity of the vehicle. Moreover, as it considers both the heading and the lateral errors, it has shown to perform well in previous studies\unskip~\cite{1131305:22465648,1131305:22465649}. Similar to the longitudinal control architecture, the lateral states of the vehicle are given as inputs ($\textbf{x}_\textbf{l} $) to the path-tracking EMRAN. The output of the Stanley controller ($\delta_s $) added to the scaled ($\textbf{K} = [\text{K}_\text{2}, \text{K}_\text{3}] $) error signal, is given as the excitation signal to the EMRAN network for learning the inverse lateral dynamics and improving the robustness.

\section{Performance Evaluation of EMRAN-aided Controller for AVs}
In this section, the performances of the proposed EMRAN-aided PID (PID-EMRAN) and Stanley (Stanley-EMRAN) controllers for a typical AV are presented for various test cases with and without external disturbances/uncertainties. As defined by the control objective in (\ref{dfg-12e413f00143}), the controllers have to minimize the tracking errors such that they rapidly converge to zero, and with minimum overshoots and steady-state errors. Moreover, the objective has to be achieved without degrading the stability of the vehicle during the maneuvers. As a first step, simulations were conducted for the longitudinal cruise control to evaluate the PID-EMRAN controller against a conventional PID method. Then, the Stanley-EMRAN controller is assessed through DLC maneuvers for lateral path-tracking. Both coupled (Coupled-EMRAN) and decoupled states are compared against a conventional Stanley controller at slow as well as high speed scenarios. Finally, quantitative comparisons of the lateral Stanley-EMRAN controller with a fuzzy logic-based method and an active disturbance rejection control scheme are shown.

Simulations have been performed in MATLAB/Simulink-Unreal Engine (UE) 4 interface on a system with Ryzen 9 5900HX processor, 32 GB of memory and Nvidia RTX 3080 GPU. This configuration was chosen to support UE visualization. The physical parameters of the vehicle used for the simulation studies are given in Table~\ref{tw-1fba46a1b919}, where $C_f $ and $C_r $ are the stiffness values of the front and rear tires, respectively. Moreover, the hyperparameters ($\epsilon_{max} $, $\epsilon_{min} $, $\gamma $, $\epsilon_{2} $, $\epsilon_{3} $, $\delta $, $N_w $, $S_w $, $\kappa $, $P_0 $, $q $, and $r $) associated with EMRAN are problem-dependent and are determined offline through an optimization by a Genetic Algorithm (GA) for achieving the best results. Note that these hyperparameters remain constant throughout the simulations. More detailed descriptions of the parameters can be obtained from\unskip~\cite{6796339, 6796942}. The GA parameters used in finding the best EMRAN parameters are as follows: crossover probability of 0.8, selection probability of 0.08, mutation probability of 0.15, maximum number of generations of 10, and a population size of 20. For a detailed description of optimization using a genetic algorithm, refer to\unskip~\cite{suresh2014hybrid}.

\begin{table}[!t]
\caption{{GA optimized parameters of the longitudinal and lateral EMRAN controllers.} }
\label{tw-a479f244405f}
\def\arraystretch{1}
\ignorespaces 
\centering 
\begin{tabular*}{\columnwidth}{@{\extracolsep{\stretch{1}}}*{3}{c}@{}}
\hline Parameters & Longitudinal & Lateral\\
\hline 
$\epsilon_{max} $ &
  7.455 &
  4.003\\
$\epsilon_{min} $ &
  3.938 &
  3.086\\
$\gamma $ &
  0.915 &
  0.981\\
$\epsilon_{2} $ &
  0.357 &
  0.005\\
$\epsilon_{3} $ &
  0.071 &
  0.003\\
$\delta $ &
  0.091 &
  0.073\\
$N_w $ &
  12 &
  9\\
$S_w $ &
  10 &
  14\\
$\kappa $ &
  0.609 &
  0.603\\
$P_0 $ &
  1.079 &
  1.155\\
$q $ &
  0.015 &
  0.001\\
$r $ &
  1.074 &
  1.120\\
\hline 
\end{tabular*}\par 
\end{table}

\begin{figure*}[!t]
    \centering
    \begin{subfigure}[t]{0.48\textwidth}
        \centering
        \includegraphics[width=\textwidth]{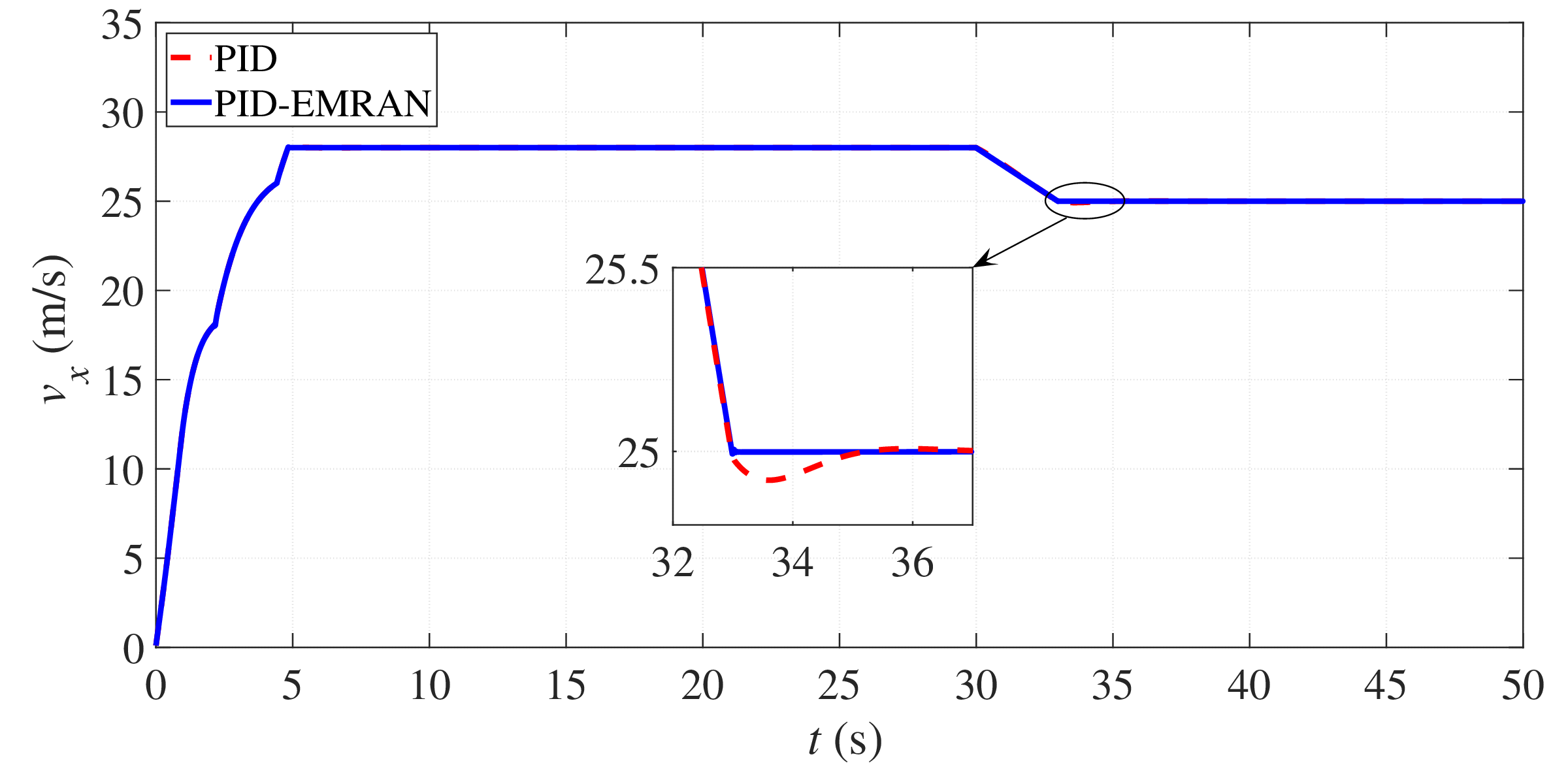}
        \caption{Cruise speed profile.}
        \label{fig-5a}
    \end{subfigure}%
    ~ 
    \begin{subfigure}[t]{0.48\textwidth}
        \centering
        \includegraphics[width=\textwidth]{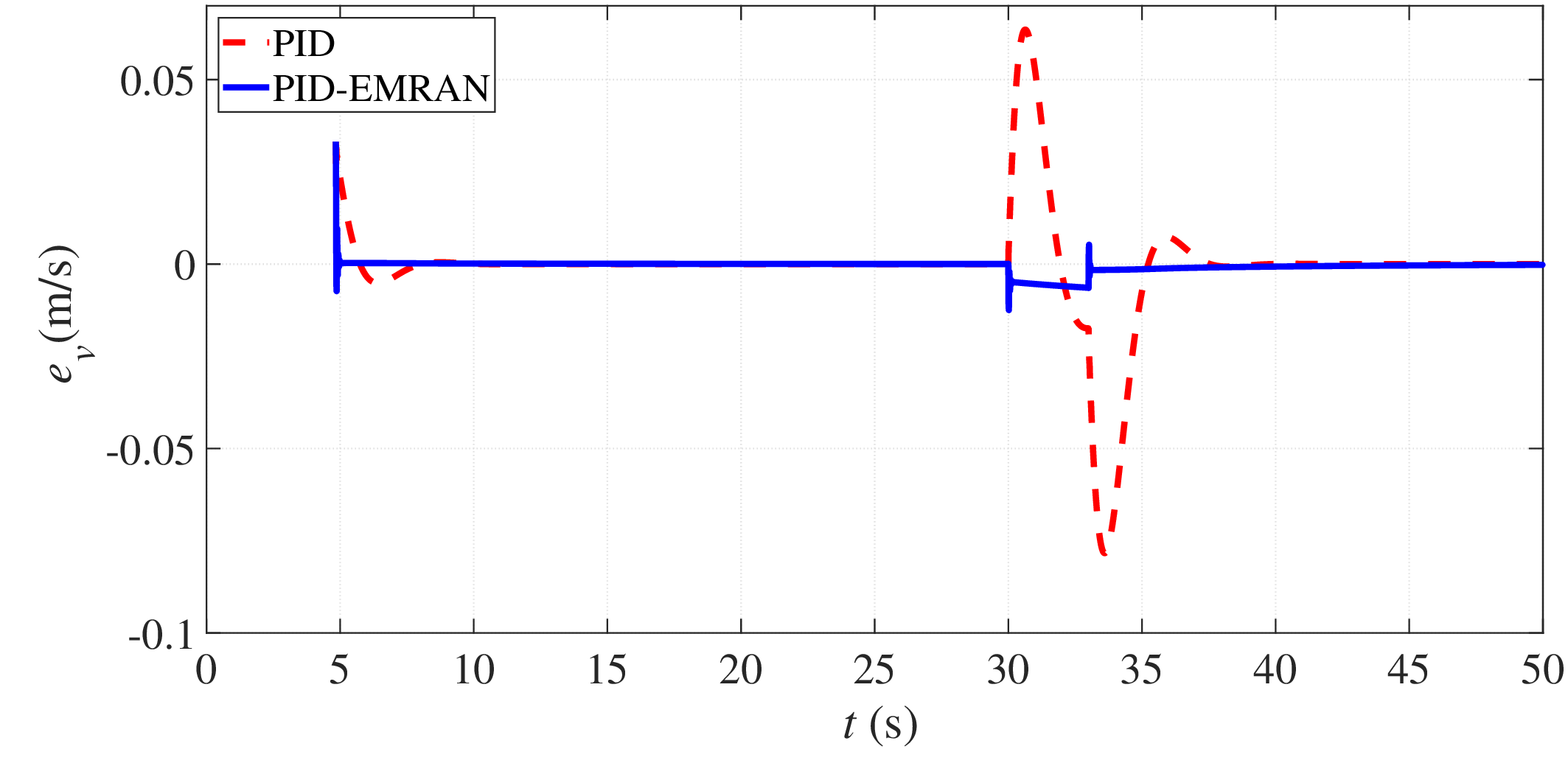}
        \caption{Velocity tracking error.}
        \label{fig-5b}
    \end{subfigure}
    \caption{Velocity tracking without disturbances and uncertainties.}
\label{f-d06d897a19f7}
\end{figure*}
For optimizing the hyperparameters associated with both the EMRAN networks, offline simulations involving speed tracking and DLC maneuvers were conducted. A reference cruise speed of 15 m/s was set for the PID aiding EMRAN to track without any disturbances and uncertainties. The optimized values were determined using GA by minimizing the objective of the fitness function, i.e., the RMS error between the reference and actual speeds. Similarly for the path-tracking EMRAN controller, a decoupled DLC maneuver at a constant velocity of 10 m/s was performed at ideal conditions, with the RMS lateral error as the objective of the fitness function to be reduced. The optimized hyperparameters of both the networks are given in Table~\ref{tw-a479f244405f}.

\begin{table}[!t]
\caption{{RMS and maximum values of cruise speed tracking errors without disturbances.} }
\label{tw-18352245d07f}
\def\arraystretch{1}
\ignorespaces 
\centering 
\begin{tabular*}{\columnwidth}{@{\extracolsep{\stretch{1}}}*{3}{c}@{}}
\hline Controller & $e_{v_{rms}} $ & $e_{v_{max}} $\\
& (m/s) & (m/s)\\
\hline 
PID &
  0.0149  &
  0.0784 \\
PID-EMRAN &
  0.0017  &
  0.0332 \\
\hline 
\end{tabular*}\par 
\end{table}

\bgroup
\begin{figure}[!b]
\centering \makeatletter\IfFileExists{images/fig7.eps}{\includegraphics[width=0.48\textwidth]{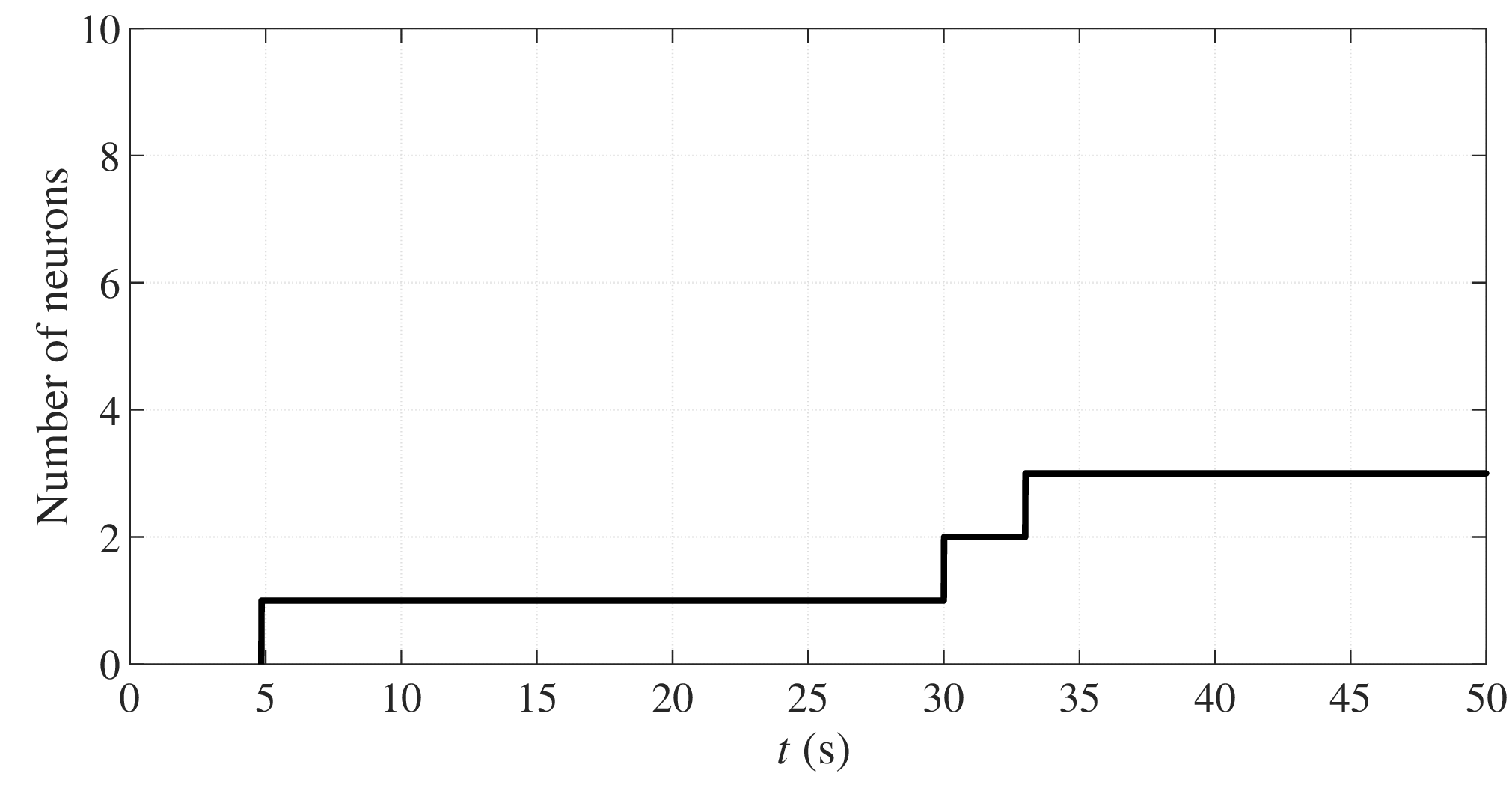}}{}
\makeatother 
\caption{{Neuron history of PID aiding EMRAN for tracking the cruise speed without any disturbances.}}
\label{f-2e2bef60f712}
\end{figure}
\egroup

\subsection{Longitudinal Cruise Control}

\subsubsection{No Disturbance}To evaluate the performance of the PID-EMRAN controller, ideal conditions without disturbances and uncertainties have been considered. The vehicle's cruising speed is changed from 28 m/s to 25 m/s at time $t = 30 $ s, and it is assumed that the vehicle travels on a dry asphalt ($\mu $ = 1.0), straight road. The proportional ($K_P $), integral ($K_I $), and derivative ($K_D $) gains associated with the baseline PID controller were tuned using GA and set as 1.841, 2.603, and 0.682, respectively.

The performances of the conventional PID and neuro-aided PID-EMRAN controllers to track the reference cruise speed are shown in Fig.~\ref{f-d06d897a19f7}. The PID-EMRAN controller outperforms the PID-based approach by lowering the overshoots and undershoots. Table~\ref{tw-18352245d07f} validates this claim, which shows the RMS ($e_{v_{rms}} $) and maximum ($e_{v_{max}} $) values of the errors. Significant improvements are observed, with the PID-EMRAN controller reducing both the $e_{v_{rms}} $ and $e_{v_{max}} $ by 88.59\% and 57.65\%, respectively. The neuron growth history in EMRAN is also shown in Fig.~\ref{f-2e2bef60f712}. The EMRAN controller adds hidden neurons to compensate for the errors during the change in cruise speed, thus tracking the reference signal more accurately during the transition.

\begin{table}[!t]
\caption{{Cruise speed tracking errors in the presence of disturbance.} }
\label{tw-28af6b930f58}
\def\arraystretch{1}
\ignorespaces 
\centering 
\begin{tabular*}{\columnwidth}{@{\extracolsep{\stretch{1}}}*{3}{c}@{}}
\hline Controller & $e_{v_{rms}} $ & $e_{v_{max}} $\\
& (m/s) & (m/s)\\
\hline
PID &
  0.1368 &
  0.5145 \\
PID-EMRAN &
  0.0163 &
  0.0817 \\
\hline
\end{tabular*}\par 
\end{table}

\begin{figure*}[!t]
    \centering
    \begin{subfigure}[t]{0.48\textwidth}
        \centering
        \includegraphics[width=\textwidth]{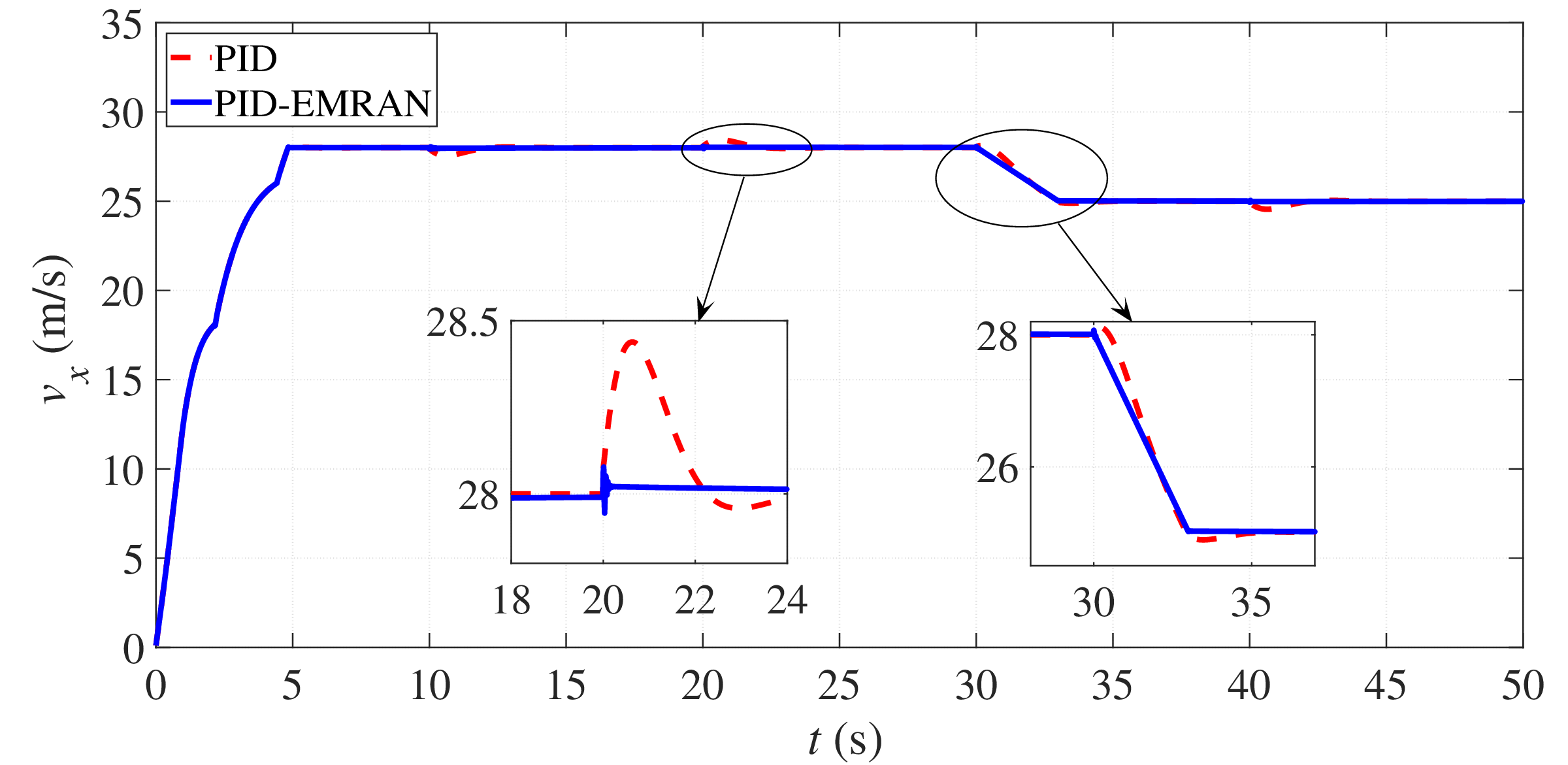}
        \caption{Cruise speed profile.}
        \label{fig-7a}
    \end{subfigure}%
    ~ 
    \begin{subfigure}[t]{0.48\textwidth}
        \centering
        \includegraphics[width=\textwidth]{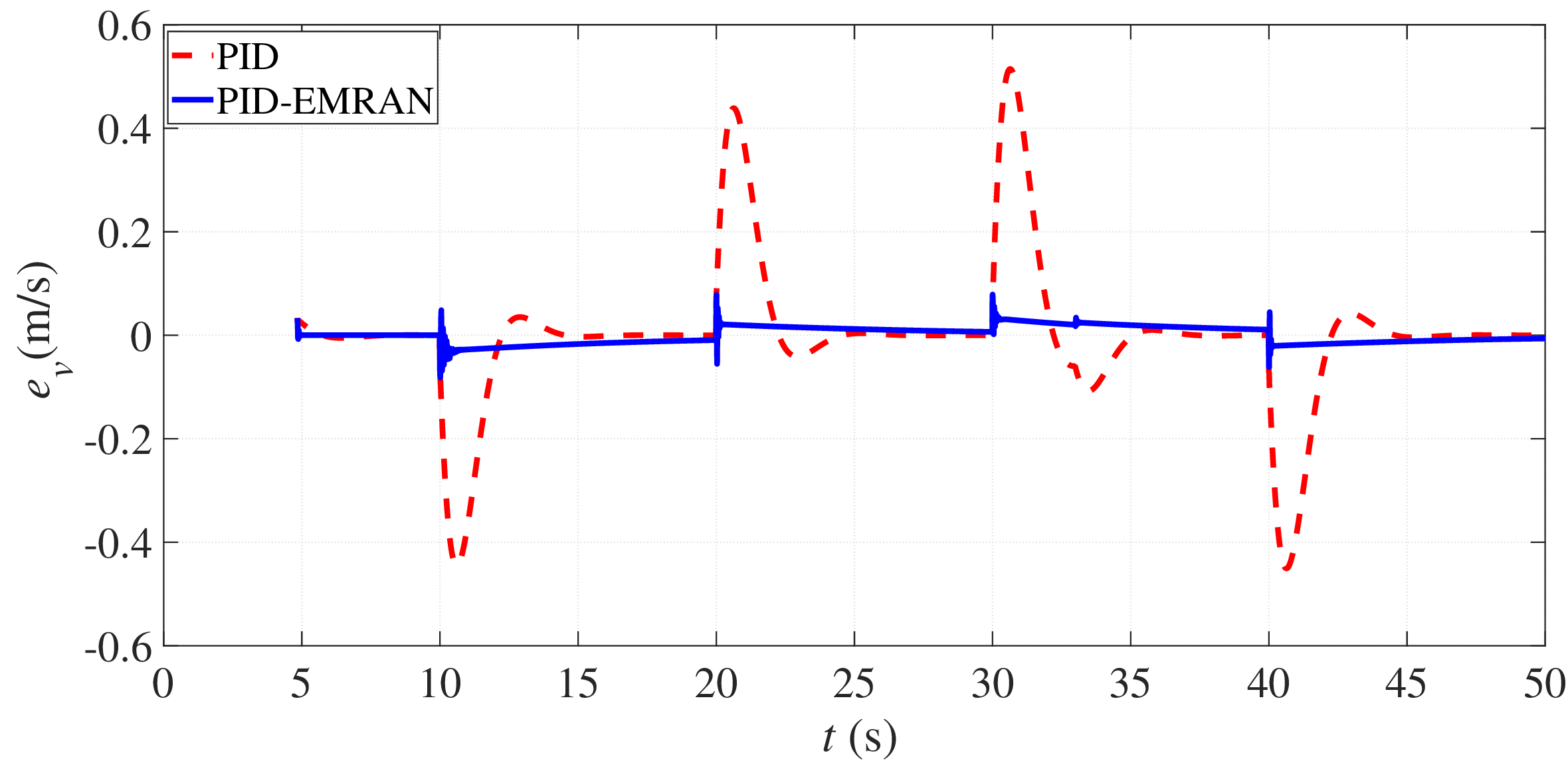}
        \caption{Velocity tracking error.}
        \label{fig-7b}
    \end{subfigure}
    \caption{Velocity tracking with varying road inclinations.}
\label{f-726cef4ce758}
\end{figure*}

\subsubsection{External Disturbances and Parametric Uncertainties}To evaluate the disturbance rejection ability of the proposed PID-EMRAN controller under extreme conditions, the vehicle was subjected to road inclinations of $\pm40\degree $. The vehicle starts on a level road initially ($t < 10 $ s) and encounters an uphill road with an inclination of $40\degree $ from $t = 10 $ s to $t = 20 $ s. At $t = 20 $ s, the slope reduces to zero, and the vehicle starts moving on the level road for another 10 seconds. Similar test cases were performed for the subsequent duration of the simulation, in which the vehicle started descending on a steep road of slope $-40\degree $ from $t = 30 $ s to $t = 40 $ s, after which it started moving on a level road again.

Figure~\ref{f-726cef4ce758} shows the performances of both the conventional PID and EMRAN-aided PID controllers in the presence of the road disturbances. The proposed control architecture shows robustness against the varying road inclinations by minimizing the overshoot/undershoots and the steady-state errors. The vehicle also maintains smooth transitions in cruise speed. It has been verified quantitatively in Table~\ref{tw-28af6b930f58}, which shows the RMS and maximum errors. For the case with disturbance, PID-EMRAN reduces the tracking errors by 86.88\% ($e_{v_{rms}} $) and 81.85\% ($e_{v_{max}} $). The neuron history is also shown in Fig.~\ref{f-24724d50d0ae}. It can be inferred by comparing Fig.~\ref{f-2e2bef60f712} and Fig.~\ref{f-24724d50d0ae} that the EMRAN controller constantly adds new neurons to mitigate the errors due to the disturbances, thus showing the effectiveness in its online learning ability. Because of the various instances of road inclinations during the simulation, no pruning of hidden neurons is observed.

Similarly, uncertainties related to vehicle parameters, ($m = m + 0.15m\sin(t) $) and ($I_y = I_y + 0.2I_y\sin(t) $) were included, where $I_y $ is the pitch moment of inertia. Additionally, environmental perturbations of the road-friction coefficient ($\mu = \mu + 0.5\mu\sin(t) $) and wind velocity ($V_w = 15 * \sin(t) $) in the longitudinal direction were also considered.

Table~\ref{tw-43a0d6f21ecd} shows that even with large parametric uncertainties and external perturbations, the proposed longitudinal controller can track the reference cruise speed with minimum errors. The PID-EMRAN scheme clearly improves the speed tracking performance with or without disturbances/uncertainties, thereby improving vehicle safety in terms of minimizing the risk of collisions with nearby vehicles.

\begin{table}[!t]
\caption{{Tracking errors with parametric uncertainties.} }
\label{tw-43a0d6f21ecd}
\def\arraystretch{1}
\ignorespaces 
\centering 
\begin{tabular*}{\columnwidth}{@{\extracolsep{\stretch{1}}}*{3}{c}@{}}
\hline Controller & $e_{v_{rms}} $ & $e_{v_{max}} $\\
& (m/s) & (m/s)\\
\hline 
PID &
  0.1463 &
  0.2678 \\
PID-EMRAN &
  0.0076 &
  0.0340 \\
\hline 
\end{tabular*}\par 
\end{table}

\bgroup
\begin{figure}[!t]
\centering \makeatletter\IfFileExists{images/fig10.eps}{\includegraphics[width=0.48\textwidth]{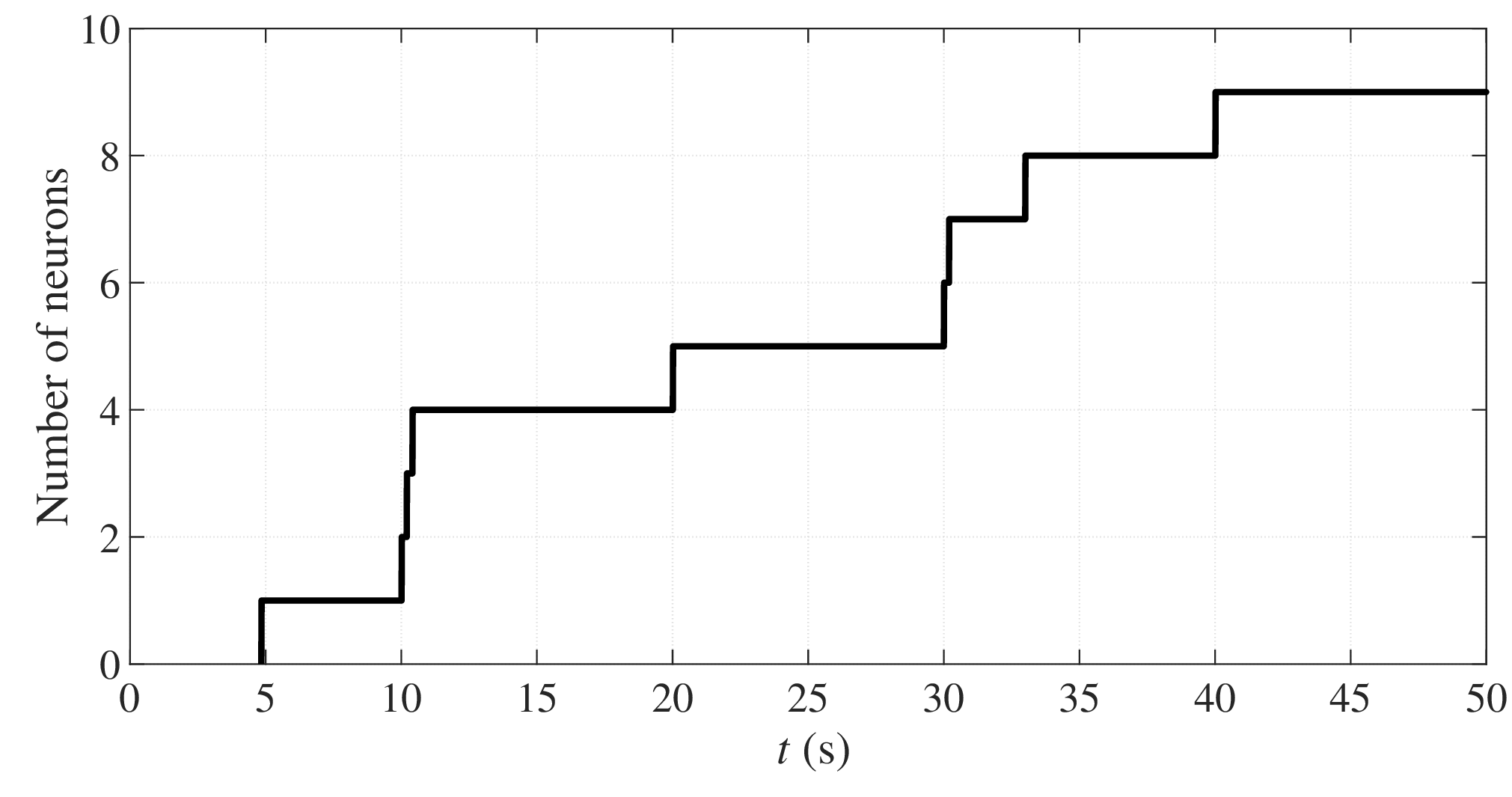}}{}
\makeatother 
\caption{{Neuron history of longitudinal EMRAN for varying road inclinations.}}
\label{f-24724d50d0ae}
\end{figure}
\egroup

\subsection{Lateral Path-tracking Control}DLC maneuvers at constant longitudinal velocities were performed to assess the lateral path-tracking ability of the proposed Stanley-EMRAN controller. The reference trajectory in terms of lateral displacement ($y_r $) and yaw angle ($\psi_r $) are expressed as in\unskip~\cite{1131305:22465637}:
\begin{align}
    \begin{split}
    &y_r = \dfrac{4.05}{2}(1+\tanh(a))-\dfrac{5.7}{2}(1+\tanh(b)) \\
    &\psi_r = \arctan\bigg(4.05\bigg(\dfrac{1}{\cosh(a)}\bigg)^{2}\bigg(\dfrac{1.2}{25}\bigg)\\
    &\quad -5.7\bigg(\dfrac{1}{\cosh(b)}\bigg)^{2}\bigg(\dfrac{1.2}{21.95}\bigg)\bigg)
    \end{split}
\tag{23}\label{dfg-72437bfe5039}\end{align}
where $a = \frac{2.4(v_xt - 27.19)}{25}-1.2 $ and $b = \frac{2.4(v_xt - 56.46)}{21.95}-1.2 $. The results of the proposed lateral controller with and without disturbances/uncertainties are presented below for both slow and high speed scenarios.

\subsubsection{Slow Speed and No Disturbance}
\begin{figure*}[!t]
    \centering
    \begin{subfigure}[t]{0.48\textwidth}
        \centering
        \includegraphics[width=\textwidth]{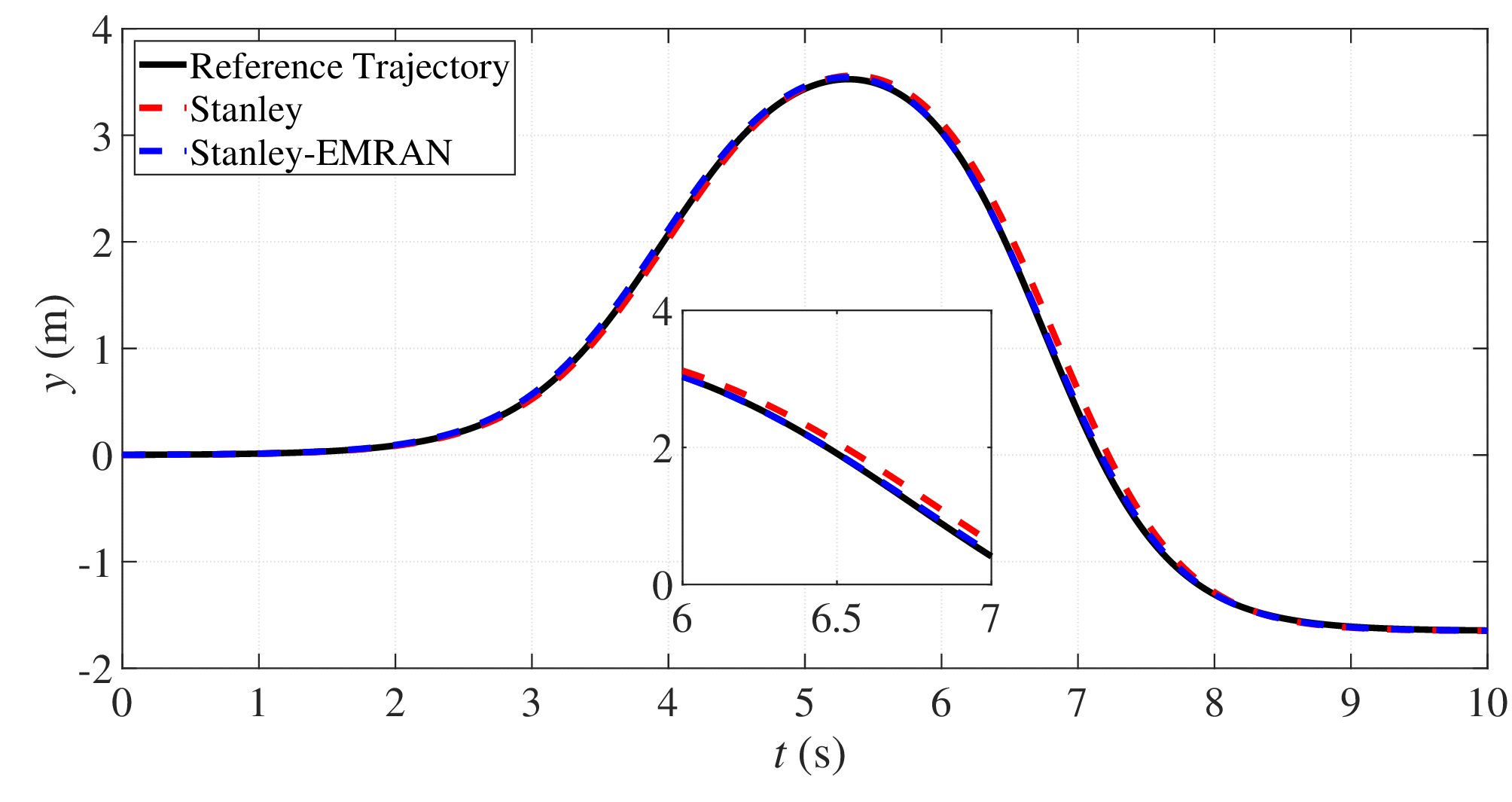}
        \caption{DLC trajectory profile.}
        \label{fig-9a}
    \end{subfigure}%
    ~ 
    \begin{subfigure}[t]{0.48\textwidth}
        \centering
        \includegraphics[width=\textwidth]{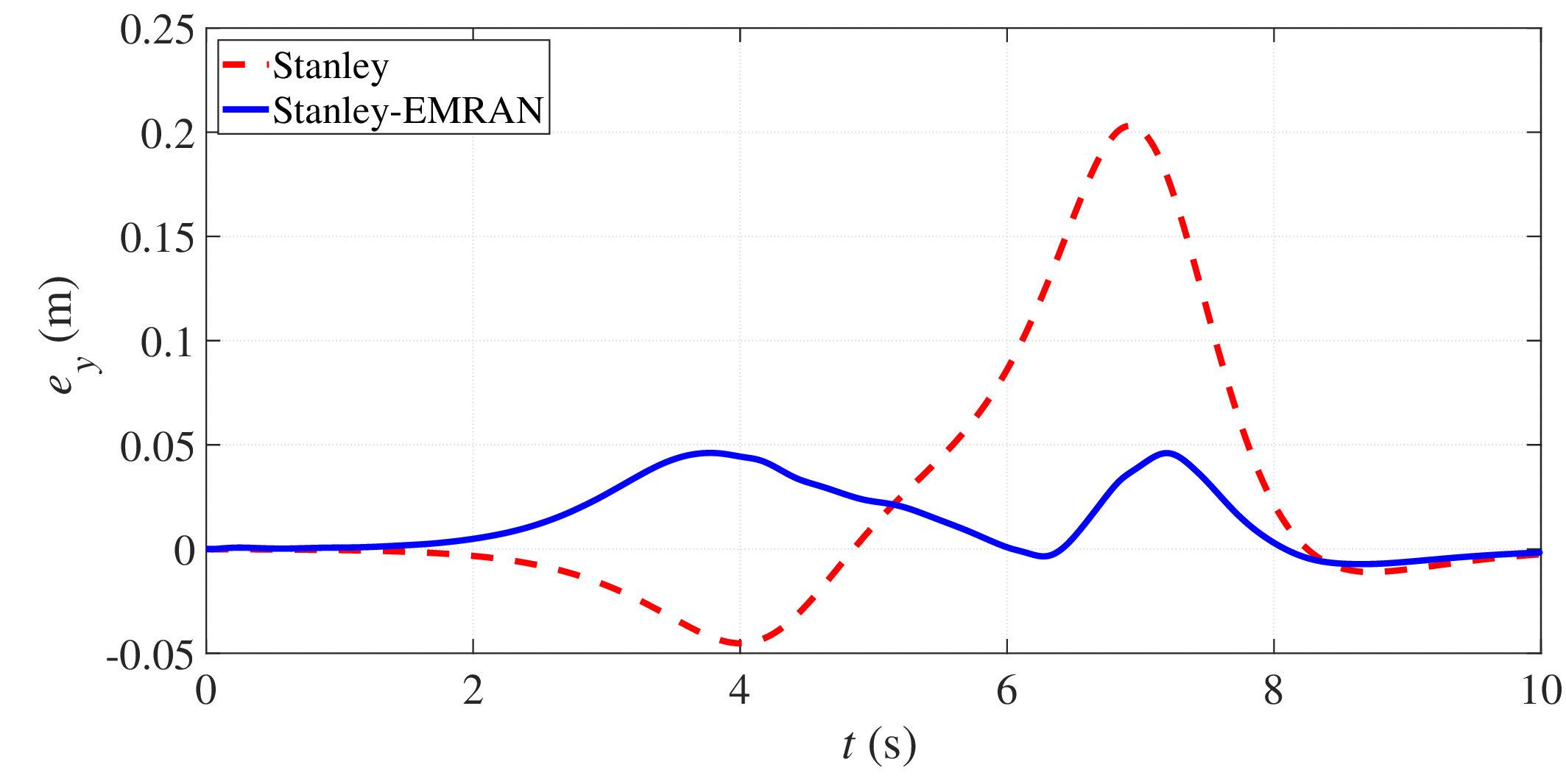}
        \caption{Lateral error.}
        \label{fig-9b}
    \end{subfigure}
    \caption{Vehicle trajectory and corresponding lateral error without external disturbances and uncertainties.}
\label{f-2153bc5d50ba}
\end{figure*}

\begin{figure*}[!t]
    \centering
    \begin{subfigure}[t]{0.48\textwidth}
        \centering
        \includegraphics[width=\textwidth]{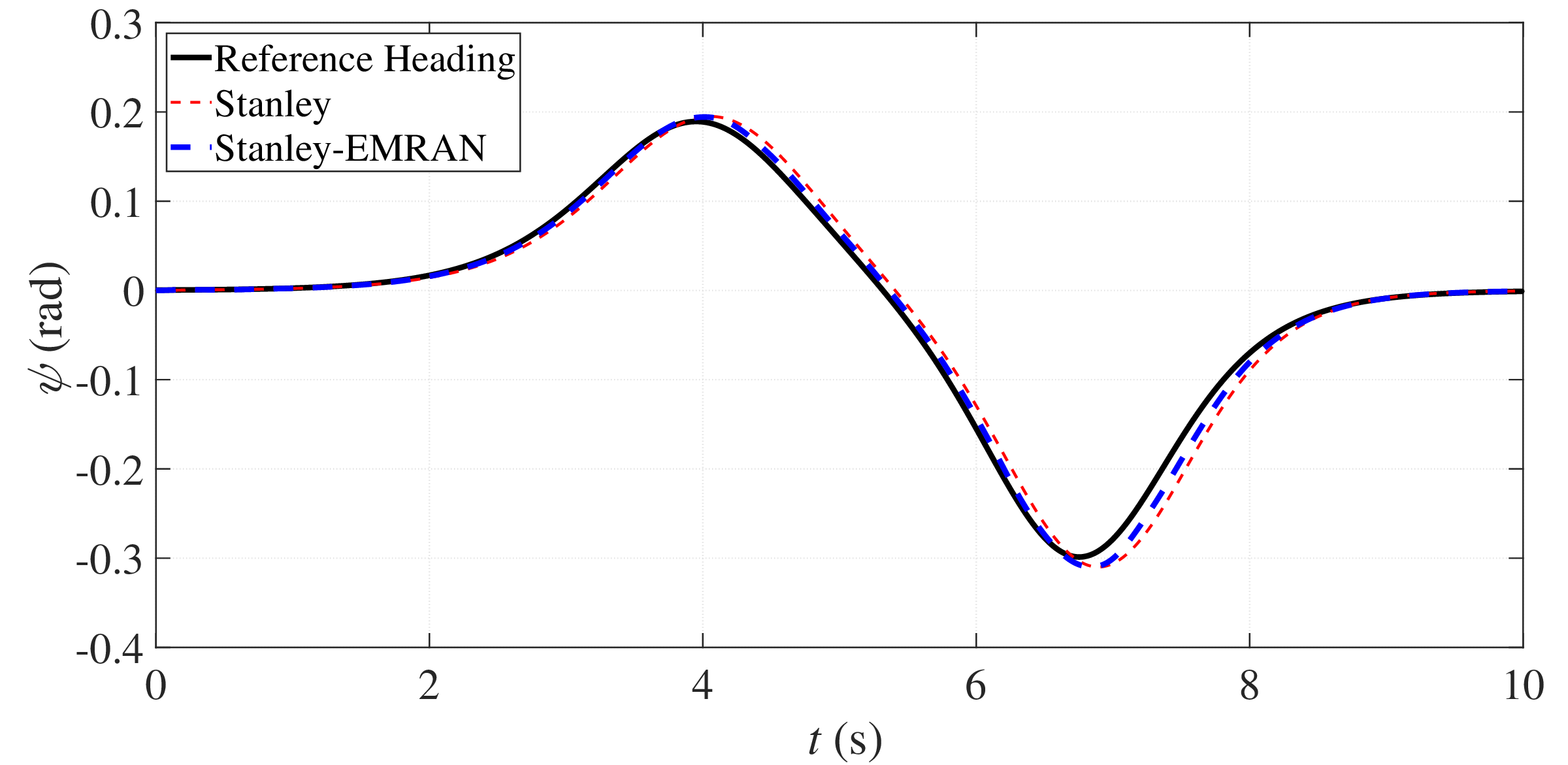}
        \caption{DLC heading profile.}
        \label{fig-10a}
    \end{subfigure}%
    ~ 
    \begin{subfigure}[t]{0.48\textwidth}
        \centering
        \includegraphics[width=\textwidth]{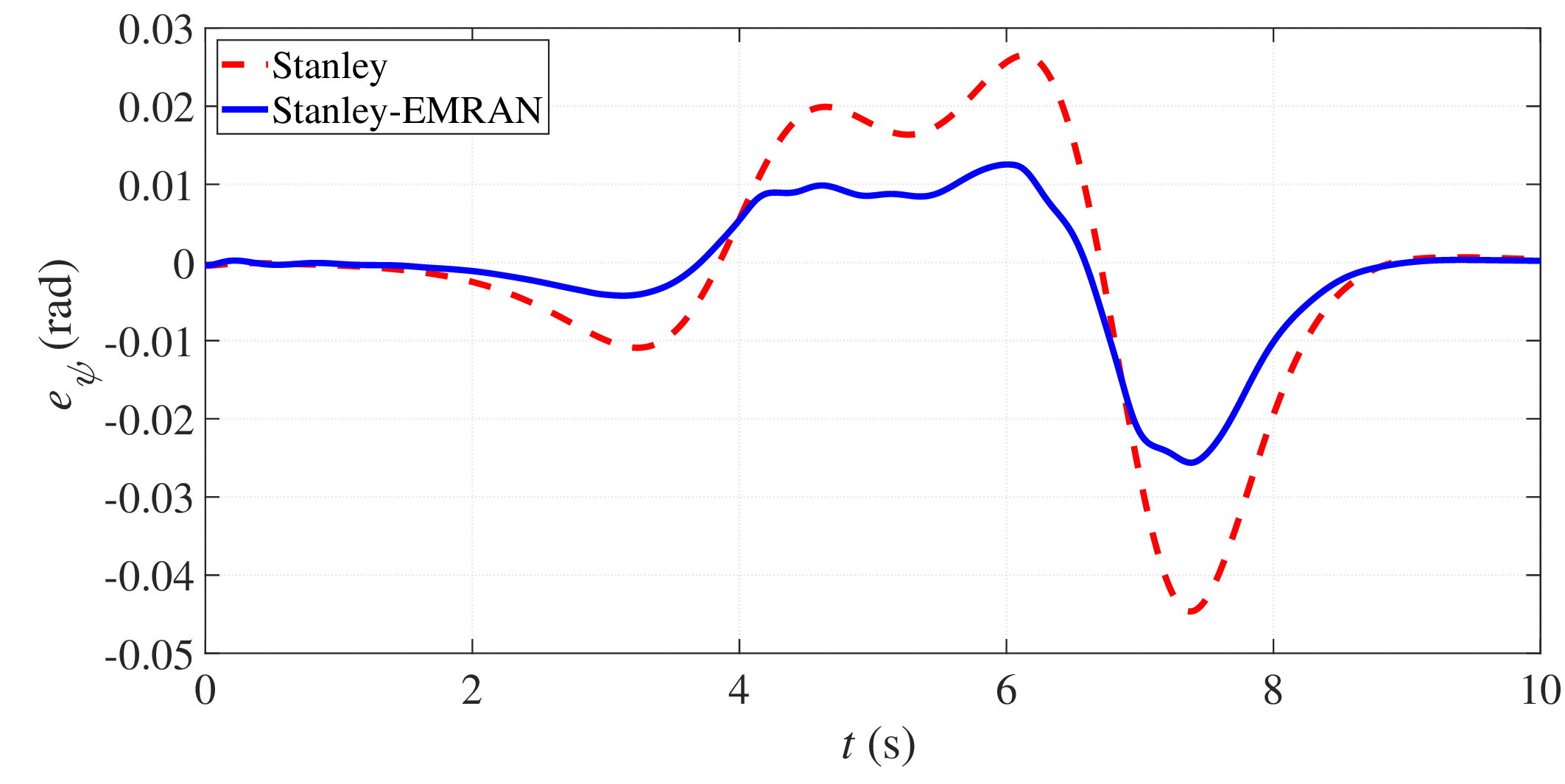}
        \caption{Heading error.}
        \label{fig-10b}
    \end{subfigure}
    \caption{Vehicle heading and error without external disturbances and uncertainties.}
\label{f-af2ee2cf8712}
\end{figure*}

This case examines the performance and online learning capability of the Stanley-EMRAN controller through a slow DLC maneuver at a constant velocity of 10 m/s, without any external disturbances and uncertainties. Figure~\ref{f-2153bc5d50ba} shows the lateral position ($y $) and the corresponding lateral error ($e_y $) for both the conventional and EMRAN-aided Stanley controllers. Table~\ref{tw-072a6f25ef22} presents a quantitative comparison of their performances in terms of the RMS and maximum values of the lateral and heading angle errors. The proposed neuro-controller improves the trajectory tracking by reducing the peak lateral offset ($e_{y_{max}} $) by 77.25\% during the DLC maneuver. The responsiveness of the vehicle to lane change has also improved compared to that of the conventional Stanley approach. Additionally, the vehicle's ability to follow the reference heading is achieved rapidly with a smaller maximum error (Fig.~\ref{f-af2ee2cf8712}), which decreased by 42.72\%. It is evident from the results that the proposed Stanley-EMRAN scheme improves the trajectory following ability of the vehicle, thereby reducing the chances of collision with another vehicle during sudden lane changes. 

\begin{table}[!t]
\caption{{Error characteristics of the lateral controllers without disturbances and uncertainties.} }
\label{tw-072a6f25ef22}
\def\arraystretch{1}
\ignorespaces 
\centering 
\begin{tabular*}{\columnwidth}{@{\extracolsep{\stretch{1}}}*{5}{c}@{}}
\hline Controller & $e_{y_{rms}} $ & $e_{y_{max}} $ & $e_{\psi_{rms}} $ & $e_{\psi_{max}} $\\
& (m) & (m) & (rad) & (rad)\\
\hline
Stanley &
  0.0683 &
  0.2031 &
  0.0160 &
  0.0447\\
Stanley-EMRAN &
  0.0218 &
  0.0462 &
  0.0089 &
  0.0256\\
Coupled-EMRAN &
  0.0274 &
  0.0677 &
  0.0083 &
  0.0267\\
\hline 
\end{tabular*}\par 
\end{table}

\begin{figure*}[!t]
    \centering
    \begin{subfigure}[t]{0.48\textwidth}
        \centering
        \includegraphics[width=\textwidth]{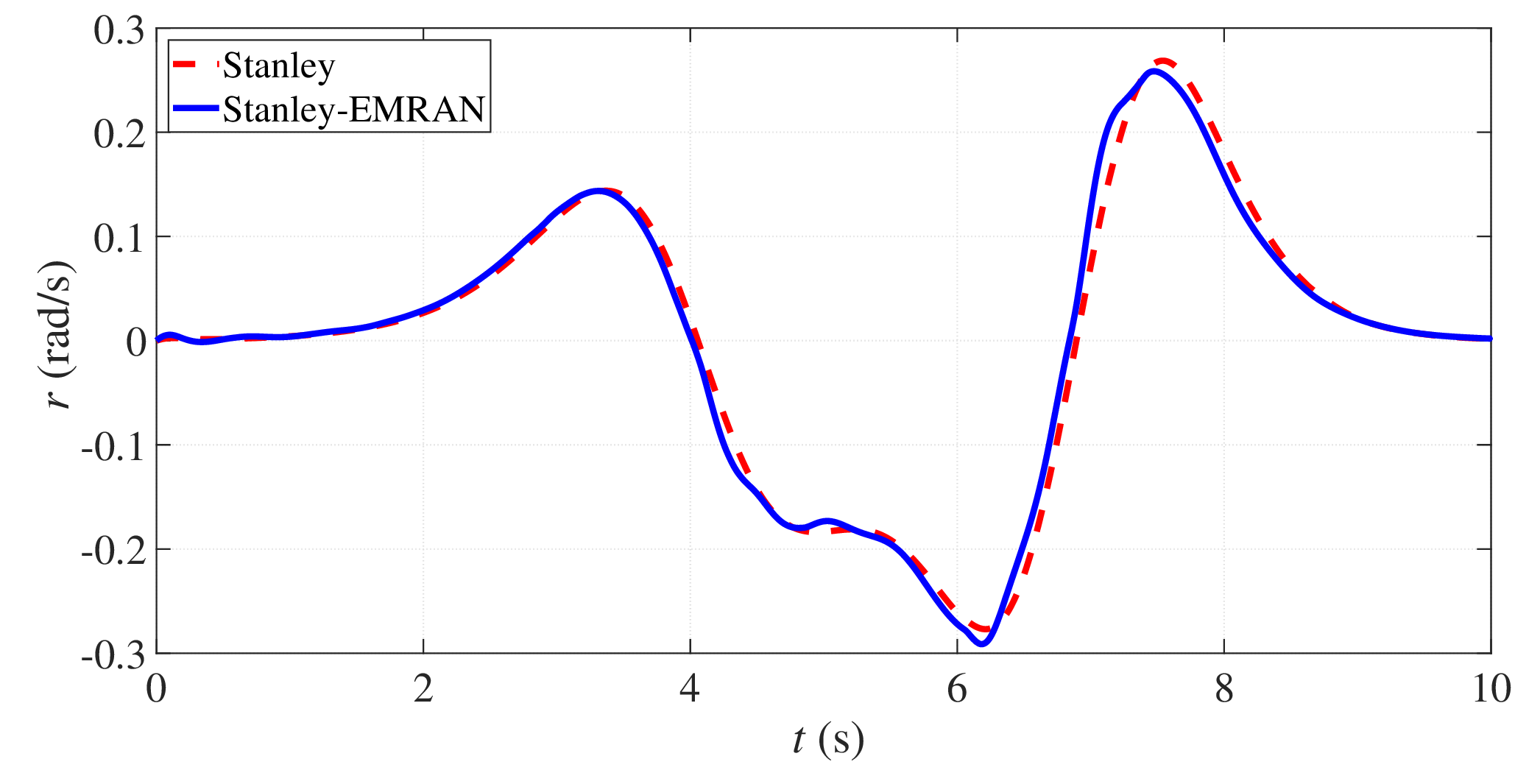}
        \caption{Yaw rate.}
        \label{fig-11a}
    \end{subfigure}%
    ~ 
    \begin{subfigure}[t]{0.48\textwidth}
        \centering
        \includegraphics[width=\textwidth]{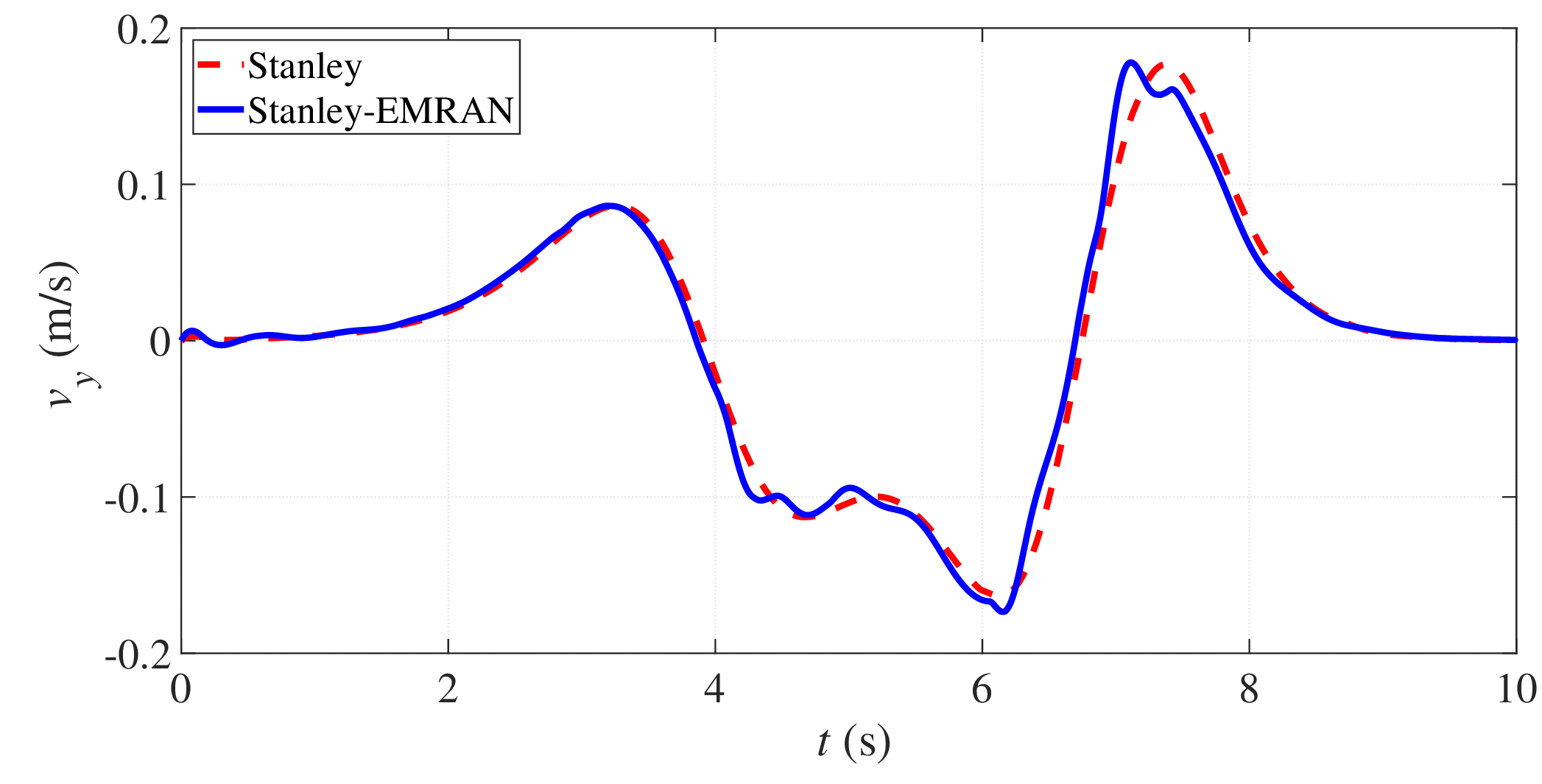}
        \caption{Lateral velocity.}
        \label{fig-11b}
    \end{subfigure}
    \caption{Lateral dynamics during the slow DLC maneuver without disturbances and uncertainties.}
\label{f-f7f058aea240}
\end{figure*}

\begin{figure*}[!t]
    \centering
    \begin{subfigure}[t]{0.48\textwidth}
        \centering
        \includegraphics[width=\textwidth]{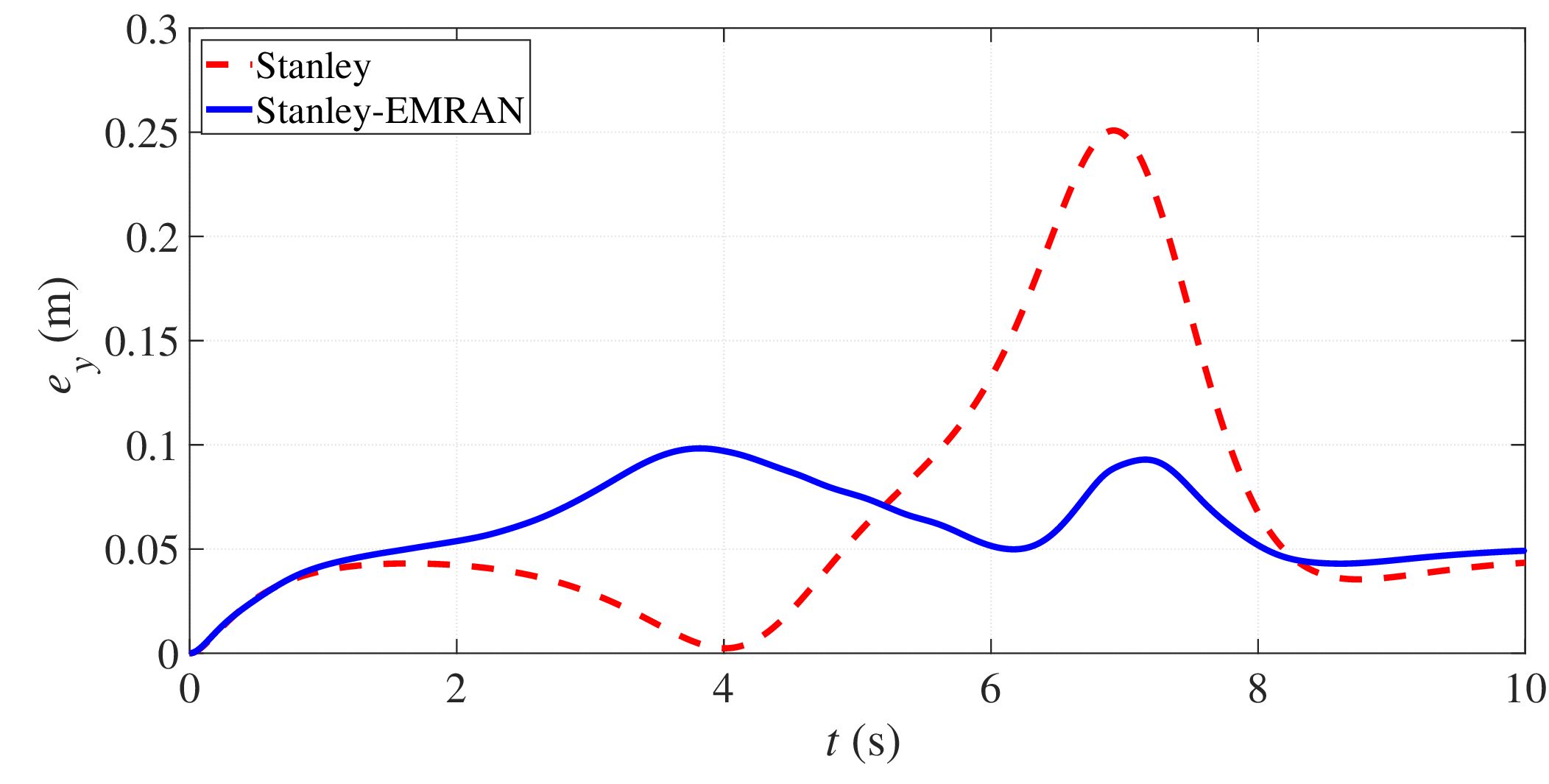}
        \caption{Lateral error.}
        \label{fig-14a}
    \end{subfigure}%
    ~ 
    \begin{subfigure}[t]{0.48\textwidth}
        \centering
        \includegraphics[width=\textwidth]{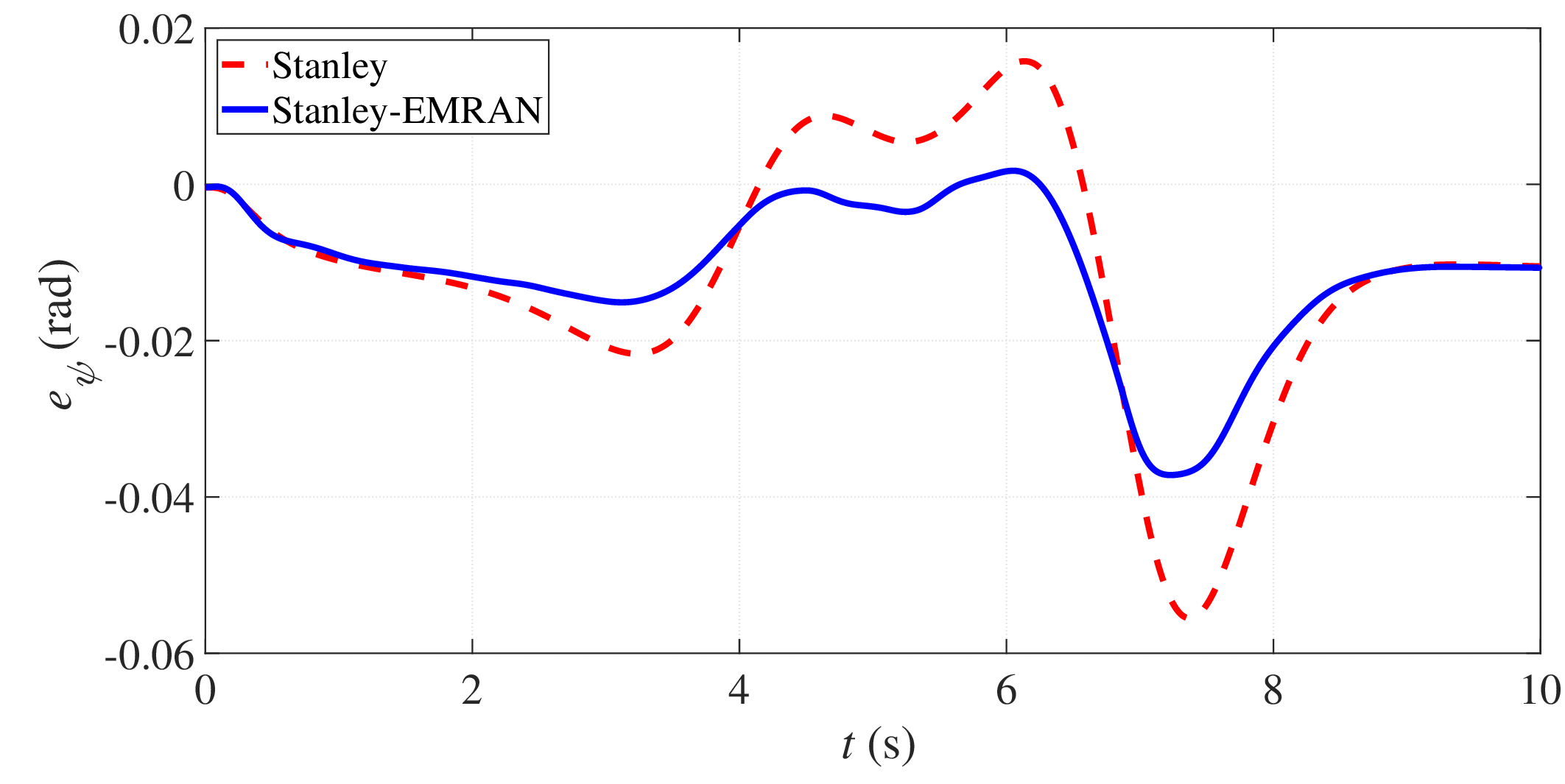}
        \caption{Heading error.}
        \label{fig-14b}
    \end{subfigure}
    \caption{Tracking errors at slow speed and in presence of a constant external force of 1500 N.}
\label{f-e91ce3816230}
\end{figure*}

\begin{table}[!t]
\caption{{Path-tracking errors at slow speed and with external disturbance.} }
\label{tw-8d703b80c8f2}
\def\arraystretch{1}
\ignorespaces 
\centering 
\begin{tabular*}{\columnwidth}{@{\extracolsep{\stretch{1}}}*{5}{c}@{}}
\hline Controller & $e_{y_{rms}} $ & $e_{y_{max}} $ & $e_{\psi_{rms}} $ & $e_{\psi_{max}} $\\
& (m) & (m) & (rad) & (rad)\\
\hline 
Stanley &
  0.0943 &
  0.2508 &
  0.0197 &
  0.0555\\
Stanley-EMRAN &
  0.0647 &
  0.0983 &
  0.0145 &
  0.0372\\
Coupled-EMRAN &
  0.0652 &
  0.1188 &
  0.0144 &
  0.0373\\
\hline 
\end{tabular*}\par 
\end{table}

\bgroup
\begin{figure}[!t]
\centering \makeatletter\IfFileExists{images/fig17.eps}{\includegraphics[width=0.48\textwidth]{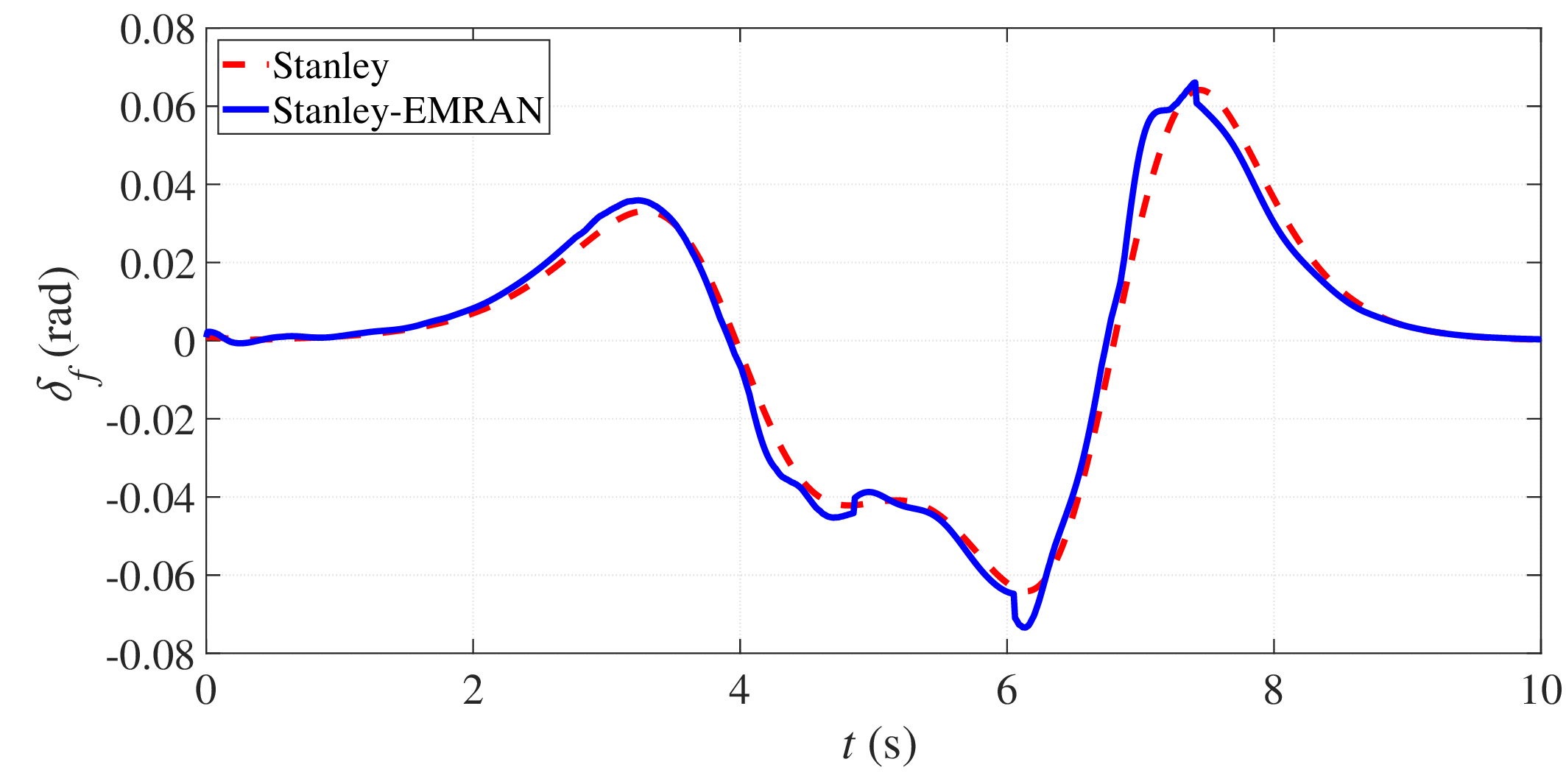}}{}
\makeatother 
\caption{{Steering input for the DLC maneuver.}}
\label{f-14b86b4f985e}
\end{figure}
\egroup

\bgroup
\begin{figure}[!t]
\centering \makeatletter\IfFileExists{images/fig18.eps}{\includegraphics[width=0.48\textwidth]{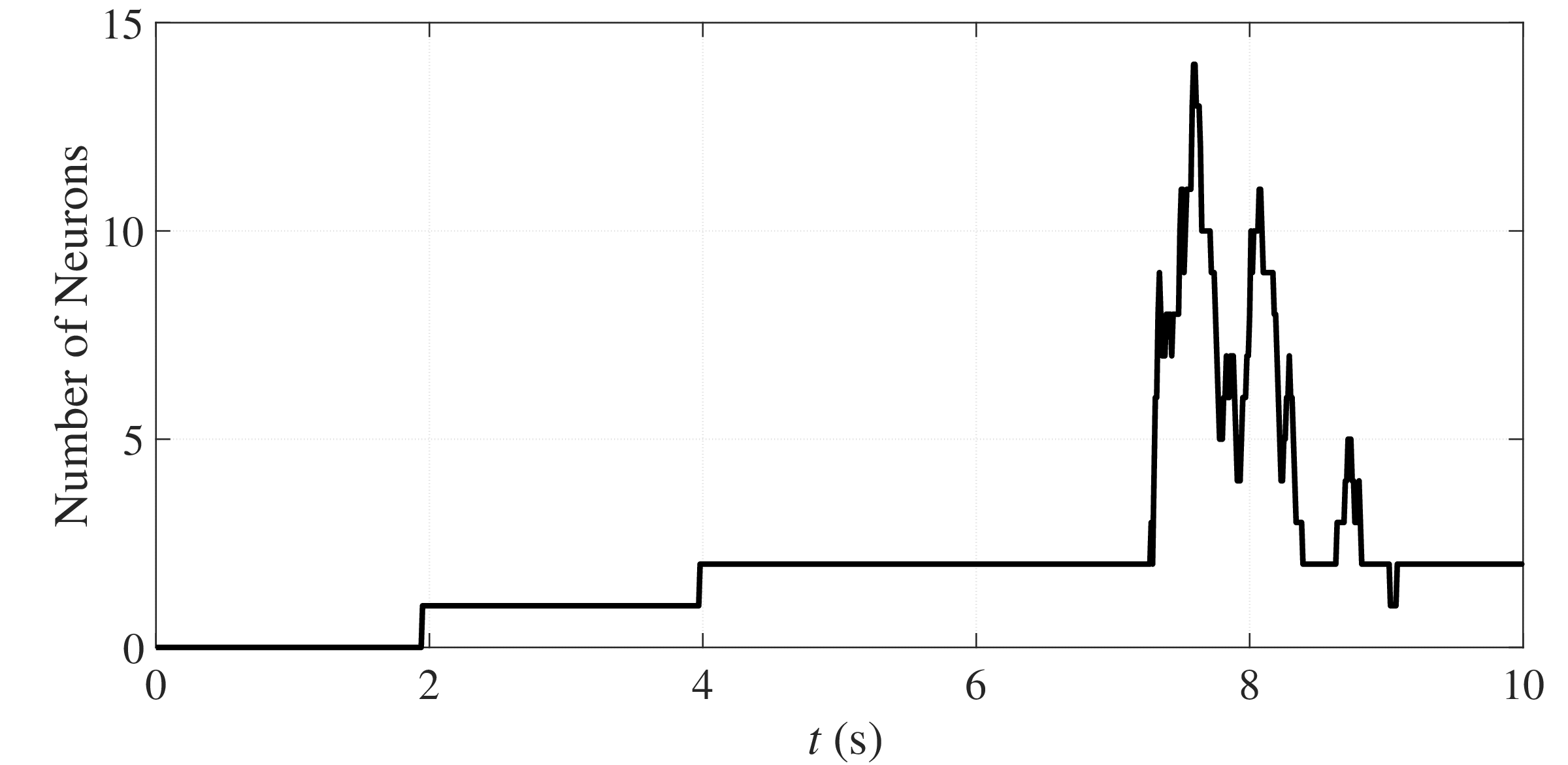}}{}
\makeatother 
\caption{{Lateral EMRAN neuron profile during the DLC maneuver at slow speed and without disturbances.}}
\label{f-087cbdd71584}
\end{figure}
\egroup

Note that the effect of longitudinal dynamics on the path-tracking performance was considered negligible during the simulations with the Stanley-EMRAN controller, and thus has limited application in practical scenarios. To address this issue, an integrated PID-EMRAN and Stanley-EMRAN controller, referred to as the Coupled-EMRAN was employed.

From Table~\ref{tw-072a6f25ef22}, it is seen that the Coupled-EMRAN controller is capable of maintaining the desired performance even under the effects of the longitudinal dynamics of the vehicle and the results of the coupled state are also comparable to that of the Stanley-EMRAN lateral control method. Minor deviations in errors are observed with the coupled architecture because of the small variations in the vehicle's velocity during the lane changes.

The yaw rate and the lateral velocity of the vehicle during the DLC maneuver are shown in Fig.~\ref{f-f7f058aea240}. Both the conventional and EMRAN-aided Stanley controllers maintain the yaw rate and the lateral velocity within a reasonable range, thereby not degrading the stability of the AV. Figure~\ref{f-14b86b4f985e} shows the control input, i.e. the steering angle ($\delta_f $) of the controllers for the DLC maneuver. The figure indicates that, at slow-speed, the vehicle is able to track the reference paths accurately, with reasonable control inputs. 

The neuron growth history is shown in Fig.~\ref{f-087cbdd71584}. It may be noted that EMRAN adds neurons when the AV is undergoing the lane changes and prunes when it is in steady-state operation. Neuron peaks are observed during $7 \,\text{s} < t < 9 \,\text{s} $ so as to minimize the large lateral and heading errors with the Stanley controller during that time period.

\begin{figure*}[!t]
    \centering
    \begin{subfigure}[t]{0.48\textwidth}
        \centering
        \includegraphics[width=\textwidth]{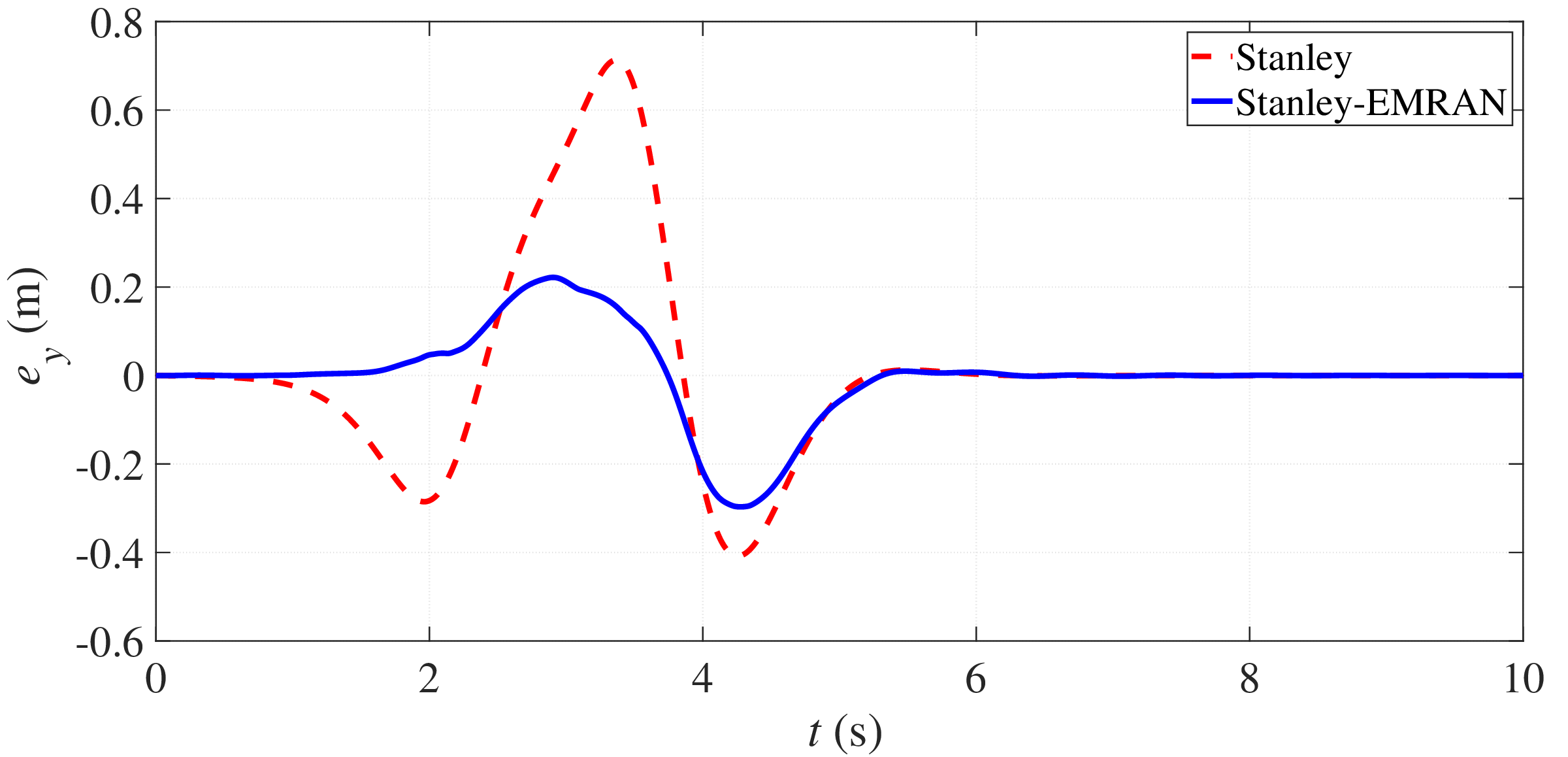}
        \caption{Lateral error.}
        \label{fig-16a}
    \end{subfigure}%
    ~ 
    \begin{subfigure}[t]{0.48\textwidth}
        \centering
        \includegraphics[width=\textwidth]{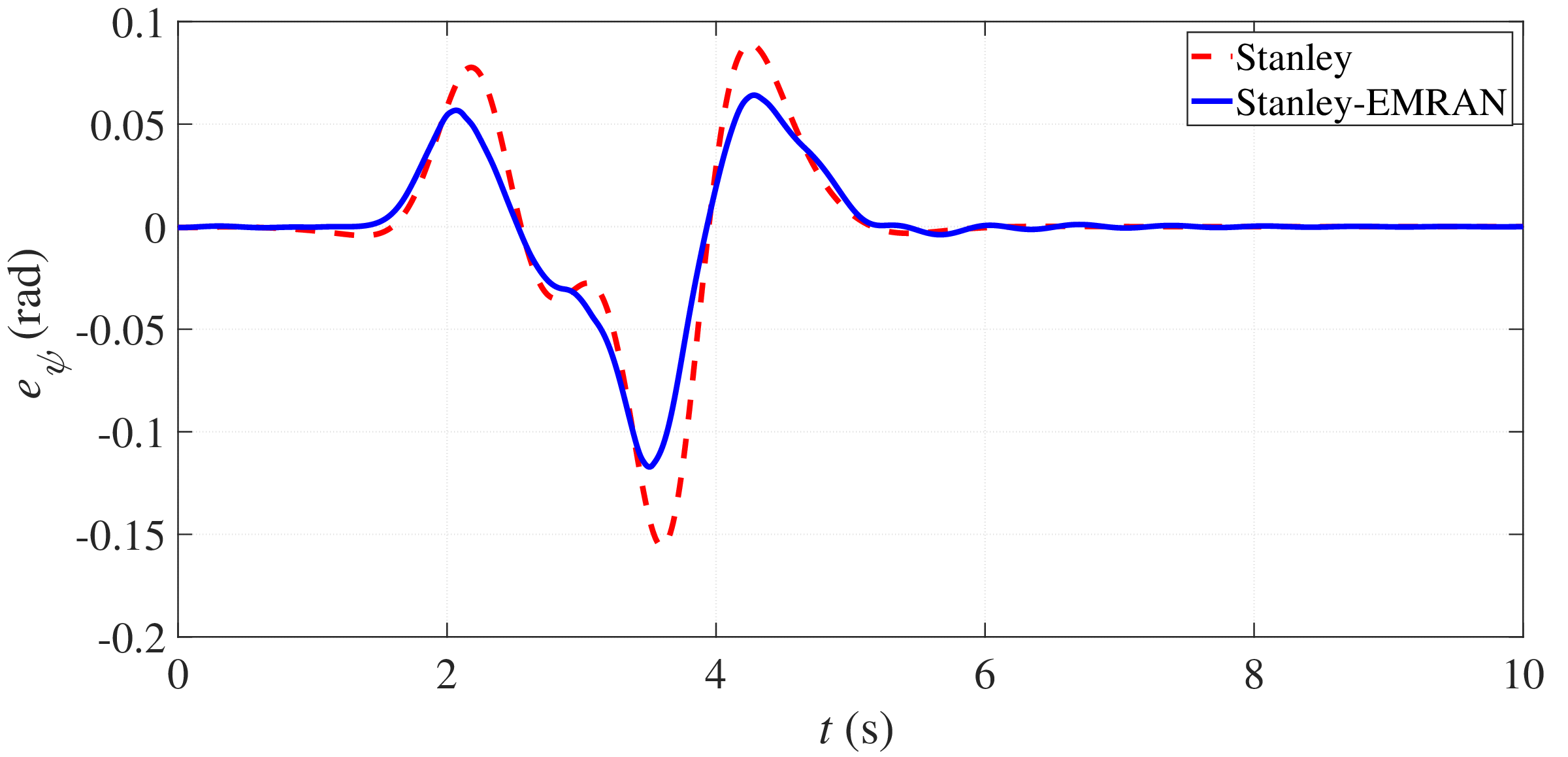}
        \caption{Heading error.}
        \label{fig-16b}
    \end{subfigure}
    \caption{Tracking errors at high speed and without disturbances.}
\label{f-228e28b857fe}
\end{figure*}

\begin{figure*}[!t]
    \centering
    \begin{subfigure}[t]{0.48\textwidth}
        \centering
        \includegraphics[width=\textwidth]{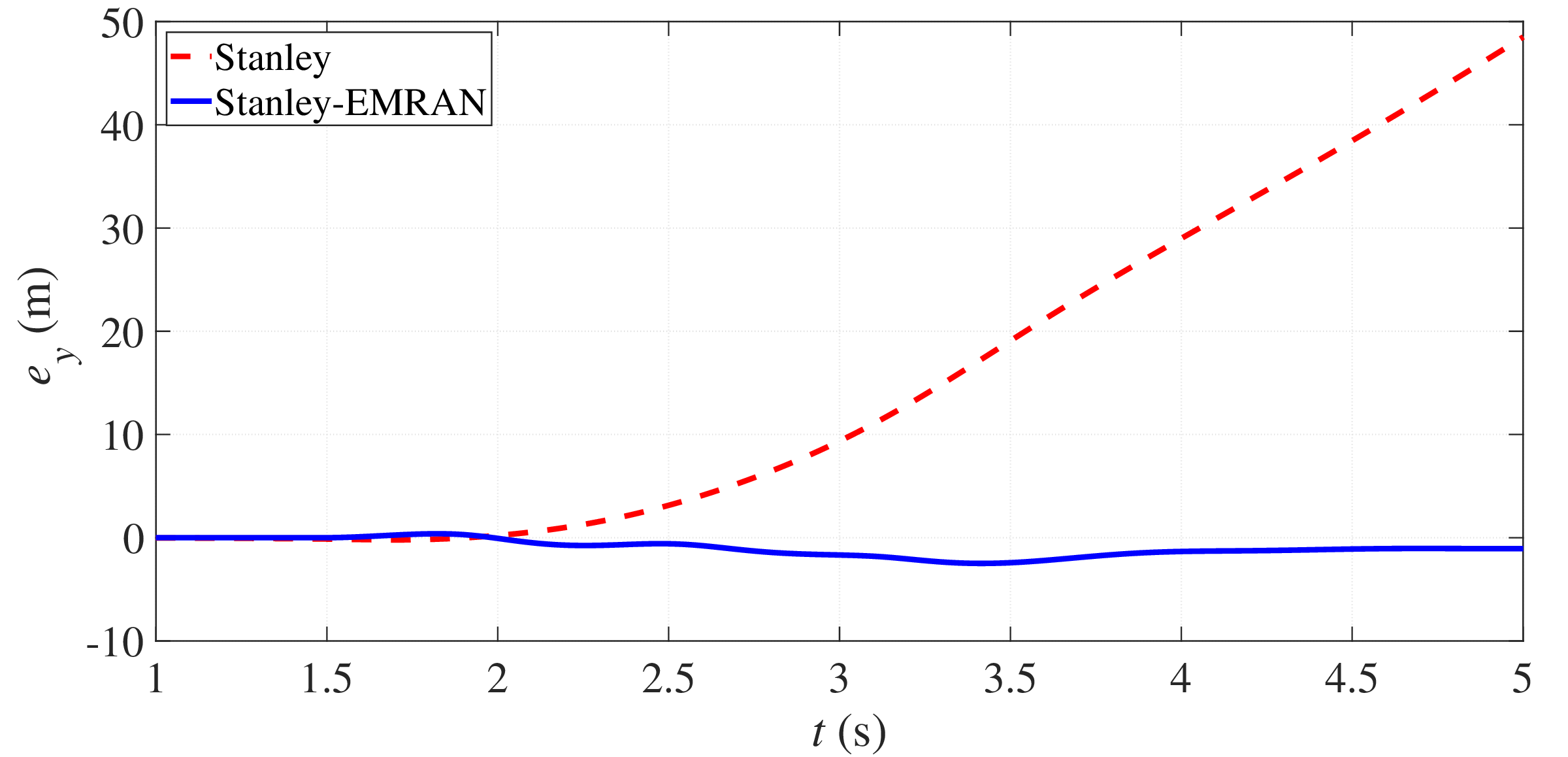}
        \caption{Lateral error.}
        \label{fig-24a}
    \end{subfigure}%
    ~ 
    \begin{subfigure}[t]{0.48\textwidth}
        \centering
        \includegraphics[width=\textwidth]{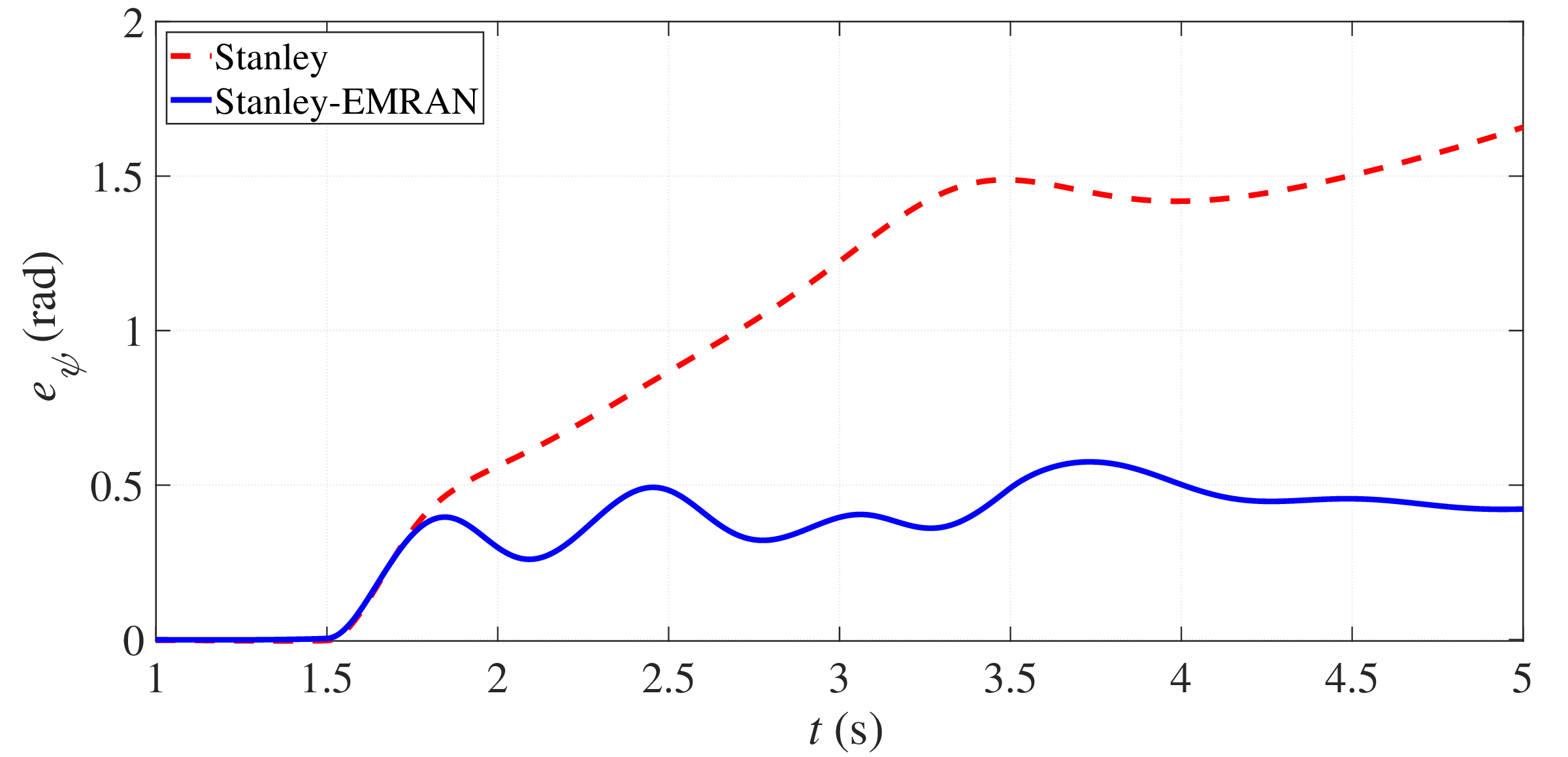}
        \caption{Heading error.}
        \label{fig-24b}
    \end{subfigure}
    \caption{Tracking errors at high speed and in the presence of wind gust.}
    \label{fig-16}
\end{figure*}

\subsubsection{Slow Speed and Disturbance}It is crucial to study the effects of external factors while designing lateral controllers for AVs. These factors can arise from wind gusts, external forces, and varying road friction in the form of disturbances and uncertainties and have the potential to degrade the desired path-tracking performance of the vehicle.

The effectiveness of the Stanley-EMRAN controller has been verified by applying a constant external lateral force of 1500 N on the vehicle. The tracking errors are shown in Fig.~\ref{f-e91ce3816230}. Initially, EMRAN starts by learning to approximate the nonlinearities due to the disturbance. During this period, the lateral errors ($e_y $) are greater compared to Stanley. It is indicative that once the learning and adaptation process is complete ($t > 4 $ s), the EMRAN-aided Stanley controller outperforms the conventional Stanley method and significantly reduces both the peaks of lateral and heading errors. Due to the bounded nature of the control input as described in (\ref{dfg-a59c7f5841b4}), the steady-state errors cannot be completely eliminated for all the cases. The bounded input ensures that the vehicle can follow the desired response set as closely as possible without losing its yaw stability. The quantitative results of the controllers are presented in Table~\ref{tw-8d703b80c8f2}. The robustness of Stanley-EMRAN to external disturbances is apparent, as it decreases $e_{y_{max}} $ by 60.8\% and $e_{\psi_{max}} $ by 32.97\%. Moreover, the results of the coupled system suggest that both the EMRAN-aided PID and Stanley controllers work coherently in rejecting the effects of the external force. 

\subsubsection{High Speed and No Disturbance}Next, we consider a high-speed DLC maneuver at constant velocity of 20 m/s and without disturbances/uncertainties. Note that it becomes more difficult for the controllers to control a fast-moving vehicle because of the shorter response time, leading to greater tracking errors.

\begin{table}[!t]
\caption{{Tracking errors at high speed and without disturbances.} }
\label{tw-2487b496cdcf}
\def\arraystretch{1}
\ignorespaces 
\centering 
\begin{tabular*}{\columnwidth}{@{\extracolsep{\stretch{1}}}*{5}{c}@{}}
\hline Controller & $e_{y_{rms}} $ & $e_{y_{max}} $ & $e_{\psi_{rms}} $ & $e_{\psi_{max}} $\\
& (m) & (m) & (rad) & (rad)\\
\hline 
Stanley &
  0.2163 &
  0.7131 &
  0.0391 &
  0.1558\\
Stanley-EMRAN &
  0.0981 &
  0.2968 &
  0.0305 &
  0.1171\\
Coupled-EMRAN &
  0.0943 &
  0.3041 &
  0.0304 &
  0.1168\\
\hline 
\end{tabular*}\par 
\end{table}

It is observed from Table~\ref{tw-2487b496cdcf} and Fig.~\ref{f-228e28b857fe} that, compared to the conventional Stanley, the proposed Stanley-EMRAN controller improves the tracking performance by minimizing the overshoots and undershoots, and peak lateral errors. However, not much improvement is noticed in the heading performance since with a single control input, both the lateral and heading errors cannot be significantly reduced simultaneously\unskip~\cite{1131305:22465617}. In this study, we have focused primarily on eliminating the lateral errors during the path-tracking maneuvers. It was found through experiments that mitigating the lateral errors while having reasonable errors in the heading, the vehicle is able to perform the harsh maneuvers without losing its stability. Aggressive heading correction also degrades the passenger comfort and increases the chance of a rollover.

Also, because of varying longitudinal dynamics of the vehicle during the lane changes, an increase in $e_{y_{max}} $ is noticed with the Coupled-EMRAN controller. This increase is also evident in all the test cases discussed earlier.

\subsubsection{High Speed and Disturbance}

\begin{table}[!t]
\caption{{Tracking errors at high speed and with disturbances.} }
\label{high-dis}
\def\arraystretch{1}
\ignorespaces 
\centering 
\begin{tabular*}{\columnwidth}{@{\extracolsep{\stretch{1}}}*{5}{c}@{}}
\hline Controller & $e_{y_{rms}} $ & $e_{y_{max}} $ & $e_{\psi_{rms}} $ & $e_{\psi_{max}} $\\
& (m) & (m) & (rad) & (rad)\\
\hline 
Stanley &
  75.3552 &
  152.1152 &
  1.3771 &
  1.6698\\
Stanley-EMRAN &
  1.0765 &
  2.4869 &
  0.3893 &
  0.5753\\
Coupled-EMRAN &
  1.0854 &
  2.5523 &
  0.3913 &
  0.5928\\
\hline 
\end{tabular*}\par 
\end{table}

A disturbance tolerance study of the conventional Stanley and EMRAN-aided Stanley controllers is discussed here. DLC maneuver at a constant vehicle speed of 20 m/s was simulated in the occurrence of a wind gust ($V_w$ = 25 m/s) at $t = 2$ s.

It is seen from Table~\ref{high-dis} and Fig.~\ref{fig-16} that the conventional Stanley controller has very poor tolerance to such an unexpected disturbance and completely diverges the vehicle from the reference path, leading to large tracking errors in both the lateral position and heading. On the other hand, the EMRAN-aided controllers were able to meet the lane change requirements and prevent the vehicle from veering off course.

\textbf{Remark: }This scenario was implemented to confirm the robustness of the proposed EMRAN-aided controllers against extreme unforeseen situations where other control approaches would fail, and should not be used as a metric for comparison.

\subsubsection{Parametric Uncertainties}Uncertainties in tire cornering stiffness ($C_f $ and $C_r $) and load ($m $ and $I_z $) are considered here at a constant speed of 10 m/s. Table~\ref{tw-ddb18bef9a33} presents the minimum and maximum values of these parameters. The RMS and maximum tracking errors with the parametric uncertainties are shown in Table~\ref{tw-c7274a7ba820}. Both the Coupled-EMRAN and Stanley-EMRAN neuro-controllers exhibit the capacity to withstand the perturbations of the internal vehicle parameters and have a better tracking accuracy compared to the conventional Stanley control.

\begin{table}[!t]
\caption{{Range of internal vehicle parameters.} }
\label{tw-ddb18bef9a33}
\def\arraystretch{1}
\ignorespaces 
\centering 
\begin{tabular*}{\columnwidth}{@{\extracolsep{\stretch{1}}}*{3}{c}@{}}
\hline Nominal value & Minimum value & Maximum value\\
\hline 
$m $ &
  $0.8m $ &
  $1.2m $\\
$I_z $ &
  $0.8I_z $ &
  $1.2I_z $\\
$C_f $ &
  $0.85C_f $ &
  $1.15C_f $\\
$C_r $ &
  $0.85C_r $ &
  $1.15C_r $\\
\hline 
\end{tabular*}\par 
\end{table}

\begin{table}[!t]
\caption{{RMS and maximum path-tracking errors with parametric uncertainties.} }
\label{tw-c7274a7ba820}
\def\arraystretch{1}
\ignorespaces 
\centering 
\begin{tabular*}{\columnwidth}{@{\extracolsep{\stretch{1}}}*{5}{c}@{}}
\hline Controller & $e_{y_{rms}} $ & $e_{y_{max}} $ & $e_{\psi_{rms}} $ & $e_{\psi_{max}} $\\
& (m) & (m) & (rad) & (rad)\\
\hline 
Stanley &
  0.0750 &
  0.2199 &
  0.0166 &
  0.0461\\
Stanley-EMRAN &
  0.0223 &
  0.0554 &
  0.0089 &
  0.0263\\
Coupled-EMRAN &
  0.0272 &
  0.0598 &
  0.0085 &
  0.0290\\
\hline 
\end{tabular*}\par 
\end{table}

\subsubsection{Comparison with other existing Methods: Type-2 Fuzzy PID and Active Disturbance Rejection Schemes}The performance of the proposed EMRAN-aided path-tracking controller is further evaluated against the results of a recently developed type-2 fuzzy PID neural network coupled to an EKF-based neural observer (EKF-T2FNN) \unskip~\cite{1131305:22465616} by using the same set of vehicle parameters and configurations. An active disturbance rejection control (ADRC) with differential flatness\unskip~\cite{1131305:22465637} is also compared with our approach.

\begin{table}[!t]
\caption{{Quantitative comparison with existing lateral controllers.} }
\label{new-case}
\def\arraystretch{1}
\ignorespaces 
\centering 
\begin{tabular*}{\columnwidth}{@{\extracolsep{\stretch{1}}}*{5}{c}@{}}
\hline Controller & $e_{y_{rms}} $ & $e_{y_{max}} $ & $e_{\psi_{rms}} $ & $e_{\psi_{max}} $\\
& (m) & (m) & (rad) & (rad)\\
\hline 
EKF-T2FNN\unskip~\cite{1131305:22465616} &
  0.0587 &
  0.0685 &
  0.0050 &
  0.0089\\
ADRC\unskip~\cite{1131305:22465616} &
  0.2207 &
  0.5593 &
  0.0178 &
  0.0363\\
Stanley-EMRAN &
  0.0218 &
  0.0462 &
  0.0089 &
  0.0256\\
\hline 
\end{tabular*}\par 
\end{table}

Table~\ref{new-case} presents the comparison of the controllers for a DLC maneuver at a constant speed of 10 m/s. The Stanley-EMRAN controller outperforms the EKF-T2FNN method in terms of minimizing the lateral $e_{y_{max}} $ and $e_{y_{rms}} $ errors by 32.5\% and 62.8\% respectively. As already stated, the primary objective of this work is to reduce the lateral errors in path-tracking without degrading the yaw stability, particularly at high speeds. This constraint leads to greater heading errors than the EKF-T2FNN controller. However, in contrast to the ADRC, the EMRAN-aided lateral controller have a better overall tracking performance.

Based on the above results, it is evident that utilizing the EMRAN neural network as an aid to feedback controllers can significantly improve the cruise control and path-tracking capabilities of an AV. Its ability to learn and adapt online makes the proposed coupled controller indispensable in achieving accurate and reliable tracking response, even in harsh and extreme conditions.

\section{Conclusions}
In this paper, a novel coupled longitudinal and lateral controller based on the online learning EMRAN neural network was presented for improving the cruise control and path-tracking performances of AVs. A feedback error learning mechanism was employed for learning the inverse dynamics of the vehicle and eliminating the effects of external disturbances and uncertainties. The performance of the controller is compared with conventional PID and Stanley approaches, as well as a fuzzy-based PID method and an active disturbance rejection control system. Simulation results in terms of the RMS and maximum tracking errors confirm the significant enhancements in control performance with the proposed scheme. In addition to providing good robustness, it is demonstrated that the EMRAN-aided controller can adapt itself under extreme situations where other controllers could fail. A self-regulated scheme integrated with the fast online learning algorithm also improves the generalization ability of the controller and significantly reduces the computational burden on the AV. There are, however, certain aspects related to the presented work that can be further investigated such as adaptive fault tolerance, actuator delays, and study of tire-road friction estimation. Future studies will examine these aspects of AV control and also include benchmarking with other neural network-based controllers through actual hardware implementation.

\bibliographystyle{IEEEtran}
\bibliography{IEEEfull}
\end{document}